\documentclass[aps,prd,preprint,preprintnumbers,showpacs]{revtex4-1}
\usepackage{graphicx}
\usepackage{amsmath}
\usepackage{caption}
\usepackage[section] {placeins}
\usepackage{slashed}
\usepackage[dvips]{color}
\usepackage{soul}

\begin{document}

\title{ Coupling vector and pseudoscalar mesons to study baryon resonances}
\author{K.~P.~Khemchandani$^1$\footnote{kanchan@rcnp.osaka-u.ac.jp}}
\author{A.~Mart\'inez~Torres$^2$\footnote{amartine@yukawa.kyoto-u.ac.jp}}
\author{H.~Kaneko$^1$\footnote{kanekoh@rcnp.osaka-u.ac.jp}}
\author{ H.~Nagahiro$^{1, 3}$\footnote{nagahiro@rcnp.osaka-u.ac.jp} }
\author{ A.~Hosaka$^1$\footnote{hosaka@rcnp.osaka-u.ac.jp}}
\preprint{YITP-11-66}

 \affiliation{
$^1$ Research Center for Nuclear Physics (RCNP), Mihogaoka 10-1, Ibaraki 567-0047, Japan.\\
$^2$ Yukawa Institute for Theoretical Physics, Kyoto University, Kyoto 606-8502, Japan.\\
$^3$ Department of Physics, Nara Women's University,  Nara 630-8506, Japan.
}

\date{\today}

\begin{abstract}
A study of meson-baryon systems with total strangeness $-1$ is made within a framework based on the chiral and hidden local symmetries. These systems  consist of octet baryons, pseudoscalar and vector mesons. The pseudoscalar meson-baryon (PB) dynamics has been earlier found determinant for the existence of some strangeness $-1$ resonances, for example, $\Lambda(1405)$, $\Lambda(1670)$, etc. The motivation of the present work is to study the effect of coupling  the closed vector meson-baryon (VB) channels to these resonances. To do this, we obtain the $PB \rightarrow PB$ and $VB \rightarrow VB$ amplitudes from the $t$-channel diagrams and the $PB \leftrightarrow VB$ amplitudes are calculated using the Kroll-Ruddermann term where, considering the vector meson dominance phenomena, the photon is replaced by a vector meson. The calculations done within this formalism reveal a very strong coupling of the VB channels to the $\Lambda(1405)$ and  $\Lambda(1670)$.  In the isospin 1 case, we find an evidence for a double pole structure of the $\Sigma (1480)$ which, like the isospin 0 resonances, is also found to couple strongly to the VB channels. 
The strong coupling  of these low-lying resonances to the VB channels can have important implications on certain reactions producing them.
\end{abstract}

\pacs{}
\maketitle

\section{Introduction}
 The dynamical generation of baryon resonances in meson-baryon systems has been studied
in detail during the past few decades using effective field theories based on chiral symmetry \cite{kaiser,baryon1,eulogiopb,baryon2,baryon3,baryon4,baryon5,baryon6,baryon7,eulogiovb,baryon8,baryon9,baryon10,baryon11,baryon12}.
In the early attempts to understand the low lying baryon resonances,
systems made of an octet pseudoscalar meson and an octet baryon  were
investigated. The resonance which has received a lot of attention in such studies is the $\Lambda(1405)$ \cite{kaiser,eulogiopb,baryon7,jido1,jido2}.
Theoretically,  the $\Lambda(1405)$  and some states in the nearby energy region are expected
to be well studied with the chiral perturbation theory involving unitarization 
 among the coupled channels.  The reason being the proximity of these resonance to the threshold of  a relevant
meson-baryon channels where the chiral perturbation theory can be used as
a good guiding principle  to determine the hadron-hadron interactions.

Some experimental investigations have been extended to further higher energy regions   where some resonances
show a large  branching ratio to channels consisting  of  vector mesons (more precisely
two or three pseudoscalar mesons with strong correlations to the vector meson
quantum numbers). 
Such states are naturally considered as candidates of dynamically
generated states with a vector meson, and several theoretical studies
have been made to verify this  with a reasonable success \cite{eulogiovb,baryon9}.
In our previous publication also, the importance 
of the vector meson-baryon interaction in   the generation of resonances
was pointed out with a more detailed formalism including $s$-, $t$-, $u$-channel diagrams 
and a contact term originating from the hidden local symmetry Lagrangian \cite{us}.
So far, most of the above studies have been carried out by
having the pseudoscalar meson-baryon (PB) and
vector meson-baryon (VB) channels independently.

Now in general, in the energy region of 1.3-2 GeV,
we naturally expect couplings between the PB and VB channels,
in particular, when the masses of the PB and VB channels are similar.
It is perhaps an aspect of the strong interaction dynamics
that we need to systematically include all possible channels.
Recently a one loop correction to the VB amplitudes has been made by
including  PB channels in the intermediate states  \cite{eobaryons10}.
The purpose of the present paper is to  carry out a full coupled channel 
calculation.

Here we 
study strangeness $-1$ baryon resonances dynamically generated from the PB and VB 
treated as coupled channels.
In the investigation dominated by S-wave interaction, as is the case of the present study,
 such a coupling appears only for 
$J^P = 1/2^-$ states, while the VB channels alone (uncoupled to PB systems)  generate both
$J^P = 1/2^-$ and $3/2^-$ states \cite{eulogiovb,us}. We will not discuss the resonances with the latter  quantum numbers
in this work since they remain unaffected by coupling PB and VB channels.
Within the framework of the low energy expansion of chiral
symmetry, PB-VB coupling is provided by an extension of the
Kroll-Ruderman (KR) term \cite{kr} for the photoproduction of a pseudoscalar meson,
by replacing the photon by a vector meson assuming the vector meson
dominance.
The  dynamics of these mesons is well described by the method
of hidden local symmetry  \cite{bando} consistently with chiral symmetry,
which is adopted in the recent studies of dynamical generation
of VB resonances.

 The basic ingredients of the present approach are, therefore,
the PB $\rightarrow$ PB amplitudes written in terms of  the Weinberg-Tomozawa interaction,
the VB $\rightarrow$ VB amplitudes   obtained from the $t$-channel vector meson exchange interaction,
and the PB $\leftrightarrow$ VB transitions acquired via the Kroll-Ruderman type coupling.
We do not include more realistic interaction of the VB channels,
because the present work is done to first inspect the kind of and the order of  
 the effect of the PB-VB coupling on the resonances well 
 understood as dynamically generated ones. 
Our study shows that considering PB and VB as coupled channels 
brings out interesting results, can generate new resonances and can be very important
in understanding the characteristics of the known resonances.
This implies that a more quantitative study  including more detailed
VB interaction should be carried out  and we should indeed make it in the near future.

We organize the paper as follows.
In section II, as a new information of the present paper,
we show briefly how the Kroll-Ruderman type
terms for the PB-VB coupling are introduced in the non-linear
sigma model.
The basic interactions in the PB and VB channels are also
briefly summarized. 
In section III, we describe the formalism to solve the Bethe-Salpeter equations
for  PB-VB coupled systems for which we use a cutoff scheme to regularize the loop integrals together with 
a form factor.
This enables us to treat the PB-VB coupling on a reasonable footing
when there is some difference in masses of PB and VB channels.
We then discuss the results of our calculations in section IV where we
show the effects of the PB-VB coupling on the states with
strangeness $S = -1$ and $J^P = 1/2^-$.
We also pay attention to the behavior of the poles of the scattering
amplitudes in the complex plane.
In section V, we summarize the findings of the present work.

\section{Interactions}
The purpose of the present paper is to study the PB-VB coupled channel interaction with the motivation to find dynamical generation
of resonances in such systems. For this purpose, it is reasonable to consider that the relative motion in the meson-baryon system is dominantly in s-wave.
The new development of this work is the inclusion of the transition between PB and VB systems in s-wave. This is done by using the KR theorem to write the Lagrangian for the $\gamma N \rightarrow \pi N$ process and by replacing the $\gamma$ by a vector meson via the  notion of the vector meson dominance. To show this procedure we start with the $\pi N$ Lagrangian from the Gell-Mann-Levi's linear sigma model,
\begin{equation}
\mathcal{L}_{\pi N} = \bar{\psi} \left[ i \gamma^\mu \partial_\mu - g_{\pi NN} \left( \sigma + i \vec{\tau}.\vec{\pi} \gamma_5 \right)  \right] \psi,\label{lpin}
\end{equation}
and define 
\begin{equation}
f U_5 = \sigma + i \vec{\tau}.\vec{\pi} \gamma_5,
\end{equation}
with
\begin{equation}
U_5 = \xi_5^2 = e^{\left(i \vec{\tau} \cdot \vec{\pi}  /f \right) \gamma_5},
\end{equation}
where $f$ is the field length
\begin{equation}
f = \left( \sigma^2 +  \vec{\pi}^2\right)^{1/2}.
\end{equation}
Further, considering the non-linear constraint 
\begin{equation}
f^2 \rightarrow f_\pi^2,
\end{equation}
where $f_\pi =$ 93 MeV is the pion decay constant, we can rewrite the Lagrangian in Eq.(\ref{lpin}) as
\begin{eqnarray}\nonumber
\mathcal{L}_{\pi N} &=& \bar{\psi} \left[ i \gamma^\mu \partial_\mu - g_{\pi NN} f_\pi \xi_5 \xi_5 \right] \psi.\\
&=& \bar{N} \xi_5^\dagger i \slashed \partial \xi_5^\dagger N - g_{\pi NN} f_\pi  \bar{N} N,\label{lpin2}
\end{eqnarray}
where to obtain the last expression we have defined $\xi_5\psi \equiv N$, $\bar{\psi} \xi_5 \equiv \bar{N}$ (which implies $\psi = \xi_5^\dagger N$, $
\bar{\psi} =\bar{N}\xi_5^\dagger$).
Subsequently, expanding  $\xi_5$ in Eq.(\ref{lpin2}) up to one pion field and introducing a vector meson field as a gauge boson of the hidden local symmetry
\begin{equation}
i \slashed \partial \longrightarrow i \slashed\partial - g \slashed \rho,
\end{equation}
we obtain
\begin{eqnarray}\nonumber
\mathcal{L}_{\pi N \rho N} &=& 
- i \frac{g }{2 f_\pi} \bar{N}   \left[ \pi ,  \rho^\mu \right]  \gamma_\mu \gamma_5 N\\
&\rightarrow& - i \frac{g g_A}{2 f_\pi} \bar{N}   \left[ \pi ,  \rho^\mu \right]  \gamma_\mu \gamma_5 N,\label{lpnrn}
\end{eqnarray}
where   $\pi = \vec{\tau} \cdot \pi$ and $\rho = \vec{\tau} \cdot \dfrac{\rho}{2}$. In the last expression above we have introduced
an arbitrary value of the nucleon axial coupling constant $g_A$, which was unity ($g_A = 1$) in the Gell-Mann-Levi's linear sigma model.
Thus, Eq.~(\ref{lpnrn}) with $g_A$ is the general Lagrangian for $\pi N \rightarrow \rho N$ to the leading order in the soft meson regime.

Next,  generalizing the Lagrangian in Eq.~(\ref{lpnrn}) for the SU(3) case, we get 
\begin{eqnarray}
\mathcal{L}_{PBVB} = \frac{-i g}{2 f_\pi} \left ( F \langle \bar{B} \gamma_\mu \gamma_5 \left[ \left[ P, V_\mu \right], B \right] \rangle + 
D \langle \bar{B} \gamma_\mu \gamma_5 \left\{ \left[ P, V_\mu \right], B \right\}  \rangle \right), \label{pbvb}
\end{eqnarray}
where the trace $\langle ... \rangle$ has to be calculated in the flavor space and $F = 0.46$, $D=0.8$ such that  $F + D \simeq  g_A = 1.26$. The ratio 
$D/(F+D) \sim 0.63$ here is close to the quark model value of 0.6, and the empirical values of $F$ and $D$ can be found, for example, in Ref.~\cite{Yamanishi:2007zza}. 

In our normalization scheme, the SU(3) matrices for the pseudoscalar ($P$) and vector mesons ($V$) are written as
\begin{eqnarray}
V =\frac{1}{2}
\left( \begin{array}{ccc}
\rho^0 + \omega & \sqrt{2}\rho^+ & \sqrt{2}K^{*^+}\\
&& \\
\sqrt{2}\rho^-& -\rho^0 + \omega & \sqrt{2}K^{*^0}\\
&&\\
\sqrt{2}K^{*^-} &\sqrt{2}\bar{K}^{*^0} & \sqrt{2} \phi 
\end{array}\right), \,\,
P =
\left( \begin{array}{ccc}
\pi^0 + \frac{1}{\sqrt{3}}\eta & \sqrt{2}\pi^+ & \sqrt{2}K^{+}\\
&& \\
\sqrt{2}\pi^-& -\pi^0 + \frac{1}{\sqrt{3}}\eta & \sqrt{2}K^{0}\\
&&\\
\sqrt{2}K^{-} &\sqrt{2}\bar{K}^{0} & \frac{-2 }{\sqrt{3}} \eta
\end{array}\right)
\end{eqnarray}
and for the baryon (B)
\begin{eqnarray}
B =
\left( \begin{array}{ccc}
 \frac{1}{\sqrt{6}} \Lambda + \frac{1}{\sqrt{2}} \Sigma^0& \Sigma^+ & p\\
&& \\
\Sigma^-&\frac{1}{\sqrt{6}} \Lambda- \frac{1}{\sqrt{2}} \Sigma^0 &n\\
&&\\
\Xi^- &\Xi^0 & -\sqrt{\frac{2}{3}} \Lambda 
\end{array}\right).
\end{eqnarray}

The Lagrangian in Eq.~(\ref{pbvb}) leads to the amplitude
\begin{eqnarray}
V^{PBVB}_{ij} = i \sqrt{3} \frac{g}{2f_\pi} C^{PBVB}_{ij}, \label{Vpbvb}
\end{eqnarray}
where, using the Kawarabayashi-Suzuki-Riazuddin-Fayazuddin relation, $g = m_\rho/\left(\sqrt{2} f_\pi \right) \sim 6$, which from now on we will denote as $g_{KR}$ (the $KR$ subscript refers to the Kroll-Ruderman coupling). To obtain this value of the coupling we have used $f_\pi = 93$~MeV and  the mass of the rho  meson  ($m_\rho = 770$ MeV). The coefficients
  $C^{PBVB}_{ij}$ in Eq.~(\ref{Vpbvb}) are given in Tables~\ref{kr_iso0} and \ref{kr_iso1}  for isospin 0 and 1, respectively.
\begin{table} [htbp]
\caption{ $C^{PBVB}_{ij}$  coefficients of the PB $\rightarrow$ VB amplitude (Eq.~(\ref{Vpbvb})) in the isospin 0 configuration.} \label{kr_iso0}
\begin{ruledtabular}
\begin{tabular}{cccccc}
&$\bar K^* N$&$\omega \Lambda$&$\rho\Sigma$&$\phi\Lambda$&$K^*\Xi$\\
\hline\\
$\bar K N$&$-D-3F$&$-\frac{1}{\sqrt{6}}(D+3F)$&$-\sqrt{\frac{3}{2}}(D-F)$&$\frac{1}{\sqrt{3}}(D+3F)$&0\\
$\pi\Sigma$&$-\sqrt{\frac{3}{2}}(D-F)$&0&$-4F$&0&$-\sqrt{\frac{3}{2}}(D+F)$\\
$\eta\Lambda$&$-\frac{1}{\sqrt{2}}(D+3F)$&0&0&0&$-\frac{1}{\sqrt{2}}(D-3F)$\\
$K\Xi$&0&$-\frac{1}{\sqrt{6}}(D-3F)$&$-\sqrt{\frac{3}{2}}(D+F)$&$\frac{1}{\sqrt{3}}(D-3F)$&$D-3F$\\
\end{tabular}
\end{ruledtabular}
\end{table}
\begin{table} [htbp]
\caption{$C^{PBVB}_{ij}$  coefficients of the PB $\rightarrow$ VB amplitude (Eq.~(\ref{Vpbvb})) in the isospin 1 configuration.}\label{kr_iso1}
\centering
\begin{ruledtabular}
\begin{tabular}{ccccccc}
&$\bar K^* N$&$\rho\Lambda$&$\rho\Sigma$&$\omega\Sigma$&$K^*\Xi$&$\phi\Sigma$\\
\hline\hline\\
$\bar K N$&$D-F$&$\frac{1}{\sqrt{6}}(D+3F)$&$-(D-F)$&$-\frac{1}{\sqrt{2}}(D-F)$&0&$D-F$\\
$\pi\Sigma$&$-(D-F)$&$\sqrt{\frac{2}{3}}\,2D$&$-2F$&0&$-(D+F)$&0\\
$\pi\Lambda$&$\frac{1}{\sqrt{6}}(D+3F)$&0&$\sqrt{\frac{2}{3}}\,2D$&0&$-\frac{1}{\sqrt{6}}(D-3F)$&0\\
$\eta\Sigma$&$-\sqrt{\frac{3}{2}}(D-F)$&0&0&0&$\sqrt{\frac{3}{2}}(D+F)$&0\\
$K\Xi$&0&$-\frac{1}{\sqrt{6}}(D-3F)$&$-(D+F)$&$\frac{1}{\sqrt{2}}(D+F)$&$-(D+F)$&$-(D+F)$\\
\end{tabular}
\end{ruledtabular}
\end{table}

To obtain the PB $\rightarrow$ PB and VB $\rightarrow$ VB amplitudes we use the $t$-channel interaction. 
The pseudoscalar-baryon Lagrangian is written as \cite{eulogiopb}
\begin{equation}
\mathcal{L}_{PB} = \langle \bar{B}i\gamma^\mu \frac{1}{4f_\pi^2} \left[ \left( \phi \partial_\mu \phi - \partial_\mu \phi \phi \right) B - B \left( \phi \partial_\mu \phi - \partial_\mu \phi \phi \right) \right]\rangle.
\end{equation}
which reduces to the amplitude of the form
\begin{equation}
V_{ij}^{PB}  = - C_{ij}^{PB}  \frac{1}{4 f_\pi^2} (K_1^0 + K_2^0).\label{Vpb}
\end{equation}
In Eqs.~(\ref{Vpb}), and throughout this article, $K_1^0$ and  $K_2^0$ refer to the energy of the meson in the initial and final state, respectively. 
Next, the VB amplitudes
 are  determined using the vector-baryon-baryon Lagrangian
\begin{eqnarray}
\mathcal{L}_{VBB}&=& -g \Biggl\{ \langle \bar{B} \gamma_\mu \left[ V^\mu, B \right] \rangle + \langle \bar{B} \gamma_\mu B \rangle  \langle  V^\mu \rangle  
\Biggr\}, \label{vbb}
\end{eqnarray}
and  the three vector-meson Lagrangian which can be obtained from 
\begin{equation}
\mathcal{L}_{3V} \in - \frac{1}{2} \langle V^{\mu\nu} V_{\mu\nu} \rangle. \label{3v}
\end{equation}
as discussed in detail in Refs.~\cite{eulogiovb,us}. As was shown in Ref.~\cite{eulogiovb}, the Lagrangians given by Eqs.~(\ref{vbb}) and (\ref{3v}) lead to the general form of the vector meson-baryon amplitude:
\begin{equation} 
V_{ij}^{VB} = - C_{ij}^{VB}  \frac{1}{4 f_\pi^2} (K_1^0 + K_2^0), \label{Vvb}
\end{equation}
which  has the structure similar to the PB $\rightarrow$ PB amplitude of Eq.~(\ref{Vpb})

The coefficients $C_{ij}^{PB}$ and $C_{ij}^{VB}$ for the pseudoscalar-baryon and  vector-baryon interactions are given in  Refs.~\cite{eulogiopb,eulogiovb}. However, for the sake of completeness we list them here again in Tables.~\ref{pb_iso0}, \ref{pb_iso1}, \ref{vb_iso0}, and \ref{vb_iso1}  for  isospins 0 and 1.
\begin{table}[htbp]
\caption{ $C_{ij}^{PB}$ coefficients of the PB $\rightarrow$ PB amplitudes ( Eq.~(\ref{Vpb})) in the isospin 0 configuration.} \label{pb_iso0}
\centering
\begin{ruledtabular}
\begin{tabular}{ccccc}
&$\bar K N$&$\pi \Sigma$&$\eta\Lambda$&$K\Xi$\\
\hline
$\bar K N$&3&$-\sqrt{\frac{3}{2}}$&$\frac{3}{\sqrt{2}}$&0\\
$\pi \Sigma$&&4&0&$\sqrt{\frac{3}{2}}$\\
$\eta\Lambda$&&&0&$-\frac{3}{\sqrt{2}}$\\
$K\Xi$&&&&3\\
\end{tabular}
\end{ruledtabular}
\end{table}

\begin{table}[htbp]
\caption{ $C_{ij}^{PB}$ coefficients of the PB $\rightarrow$ PB amplitudes  (Eq.~(\ref{Vpb})) in the isospin 1 configuration.} \label{pb_iso1}
\centering
\begin{ruledtabular}
\begin{tabular}{cccccc}
&$\bar K N$&$\pi\Sigma$&$\pi\Lambda$&$\eta\Sigma$&$K\Xi$\\
\hline
$\bar K N$&1&$-1$&$-\sqrt{\frac{3}{2}}$&$-\sqrt{\frac{3}{2}}$&0\\
$\pi\Sigma$&&2&0&0&1\\
$\pi\Lambda$&&&0&0&$-\sqrt{\frac{3}{2}}$\\
$\eta\Sigma$&&&&0&$-\sqrt{\frac{3}{2}}$\\
$K\Xi$&&&&&1\\
\end{tabular}
\end{ruledtabular}
\end{table}

\begin{table}[htbp]
\caption{$C_{ij}^{VB}$ coefficients of the VB $\rightarrow$ VB amplitudes  (Eq.~(\ref{Vvb})) in the isospin 0 configuration.}\label{vb_iso0}
\centering
\begin{ruledtabular}
\begin{tabular}{cccccc}
&$\bar K^* N$&$\omega \Lambda$&$\rho\Sigma$&$\phi\Lambda$&$K^*\Xi$\\
\hline
$\bar K^* N$&3&$\sqrt{\frac{3}{2}}$&$-\sqrt{\frac{3}{2}}$&$-\sqrt{3}$&0\\
$\omega\Lambda$&&0&0&0&$-\sqrt{\frac{3}{2}}$\\
$\rho\Sigma$&&&4&0&$\sqrt{\frac{3}{2}}$\\
$\phi\Lambda$&&&&0&$\sqrt{3}$\\
$K^*\Xi$&&&&&3\\
\end{tabular}
\end{ruledtabular}
\end{table}

\begin{table}[htbp]
\caption{$C_{ij}^{VB}$ coefficients of the VB $\rightarrow$ VB amplitudes ( Eq.~(\ref{Vvb})) in the isospin 1 configuration.}\label{vb_iso1}
\centering
\begin{ruledtabular}
\begin{tabular}{ccccccc}
&$\bar K^* N$&$\rho\Lambda$&$\rho\Sigma$&$\omega\Sigma$&$K^*\Xi$&$\phi\Sigma$\\
\hline
$\bar K^* N$&1&$-\sqrt{\frac{3}{2}}$&$-1$&$-\frac{1}{\sqrt{2}}$&0&1\\
$\rho\Lambda$&&0&0&0&$-\sqrt{\frac{3}{2}}$&0\\
$\rho\Sigma$&&&2&0&1&0\\
$\omega\Sigma$&&&&0&$-\frac{1}{\sqrt{2}}$&0\\
$K^*\Xi$&&&&&1&1\\
$\phi\Sigma$&&&&&&0\\
\end{tabular}
\end{ruledtabular}
\end{table}

It should be mentioned at this point that it has been shown in Ref.~\cite{us} that the contributions obtained from the  $u$-channel and a contact interaction coming from the hidden gauge Lagrangian together with the $t$-channel 
interactions are important in the study of  VB systems. In view of the findings of Ref.~\cite{us}, in principle, considering the  $t$-channel exchange alone would lead to an incomplete information. However, firstly, the purpose of the present work is to study the properties of strangeness $-1$ low energy resonances, like $\Lambda (1405)$, when
the  heavier VB  channels are coupled to the PB systems. And secondly, we cannot {\it a priori} know if the PB-VB coupling would bring important new information. We, thus, start with a simplified model where  the VB $\rightarrow$ VB amplitudes are obtained from the $t$-channel,  while neglecting other contributions as done in a previous study of the VB systems \cite{eulogiovb}.

We would like to point out that the VB interaction (~Eq.~(\ref{Vvb})~) determined from the $t$-channel exchange gives rise to  spin independent amplitudes. Thus, the amplitudes obtained and the poles found in the VB system possess spin 1/2 and 3/2. However, the coupling of the PB to VB channels  in s-wave, as is the case of the present study, would affect only the spin 1/2 amplitudes.  Consequently, only the results obtained for spin 1/2 meson-baryon systems will be discussed in the present work.

Finally,  the whole set of interactions can be summarized as shown in  Table~\ref{allint}. 

\begin{table}[htbp]
\caption{A summary of all the interactions.}\label{allint}
\centering
\begin{ruledtabular}
\begin{tabular}{ccc}
$V_{ij}$ & VB&PB\\
\hline\hline\\
VB&  $- C_{ij}^{VB}  \dfrac{1}{4 f_\pi^2} (K_1^0 + K_2^0)$&$-i \sqrt{3} \dfrac{g_{KR}}{2f_\pi} C^{PBVB}_{ij}$ \\
&&\\
PB&$i \sqrt{3} \dfrac{g_{KR}}{2f_\pi} C^{PBVB}_{ij}$& $- C_{ij}^{PB}  \dfrac{1}{4 f_\pi^2} (K_1^0 + K_2^0)$\\
\end{tabular}
\end{ruledtabular}
\end{table}

\section{Formalism}
\subsection{Solving coupled channel equations}
In the previous section we have discussed the basic interactions for PB-VB coupled systems. Using the tree level amplitudes, summarized in Table~ \ref{allint},  as the kernels  $V$,  we solve  the Bethe-Salpeter equation
\begin{equation}
T = V + V G T, \label{bs}
\end{equation}
which, for a  single meson-baryon channel case,  can be explicitly written as 
\begin{equation}
T  (k_1, k_2) =  V (k_1, k_2) + 2 M i \int \frac{d^4q}{(2\pi)^4}   \frac{  V(k_1, q) T(q, k_2)}{\left( q^2 - m^2 + i\epsilon \right) \left(\left( P - q \right)^2  -M^2 + i\epsilon\right)}, \label{bsintegral}
\end{equation}
where  $k_1 (k_2)$ is the four momentum of the  meson in the initial (final) state, $m,\, M $ (here and throughout this article) are the meson and baryon masses, respectively, and $P$ is the total four momentum of the system. It has been shown in earlier works (for example, in Refs.\cite{eulogiopb, oller}) that in the low energy studies of two hadron systems based on chiral unitary dynamics, $V$ and $T$ can be factorized out of the loop integral in Eq.~(\ref{bsintegral}). This simplifies the integral Bethe-Salpeter equation to an algebraic one since the calculation of the loop integral involves only  the   $G$ function
\begin{equation}
G  =   2 M i \int \frac{d^4q}{(2\pi)^4}   \frac{ 1}{\left( q^2 - m^2 + i\epsilon \right) \left(\left( P - q \right)^2  -M^2 + i\epsilon\right)}. \label{loopfn}
\end{equation}
These loop functions are divergent in nature and one either
calculates them by using the dimensional regularization method or the cut-off method.  Since our calculations involve a larger energy range  (as compared to the previous studies of the uncoupled PB-VB systems), we first study  the behavior of the loops calculated by using these two different regularization schemes to verify if they work well far from the threshold of the meson-baryon systems considered here. In doing so, we encountered a problem with the loop calculated using the dimensional regularization method. It is worth discussing this issue in some detail before going ahead. 

\subsection{Calculation of the loops}
The aim of the present work is to  study the effect of  the  VB channels on the poles found in the PB systems. From the time independent perturbation theory, one would expect, up to the second order, a correction to the mass of a resonance  to be proportional to
\begin{equation}
\Delta E  \propto \sum\limits_q   \frac{\mid \langle VB \mid V \mid PB \rangle\mid^2} {E - E(q)}, \label{pertE}
\end{equation}
where $E$ is the total energy of the unperturbed system, and $E(q)$ is the energy of an intermediate state labeled by $q$. The contribution of the above equation should be negative for the closed channels, for which $E(q) > E$, and one would n\"aively expect a pole to shift to lower energies if a coupling to a closed channel is introduced. 
Similarly, from the definition of the loop function, $G$ (Eq.~(\ref{loopfn})), one would expect  the loop function to possess certain features, like its imaginary part must be zero at energies below the threshold ($\sqrt{s} <  m + M$) while its real part should be negative.  Let us now look at the loop function calculated with different regularizing schemes: the cut-off and the dimensional regularization.

\begin{figure}[h]
\begin{minipage}[b]{0.49\linewidth}
\centering
\includegraphics[width=8cm,height=7cm]{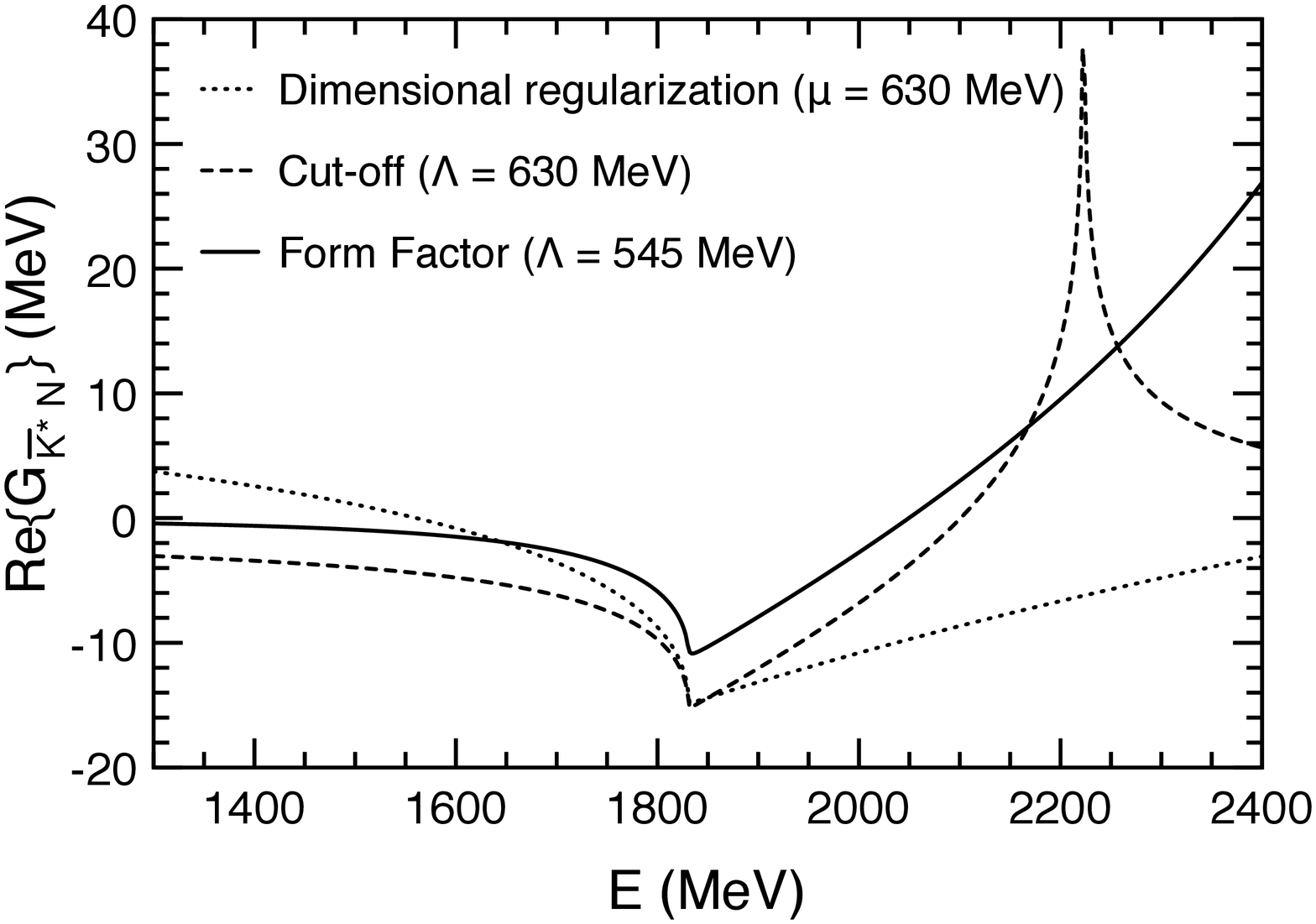}
\end{minipage}
\begin{minipage}[b]{0.49\linewidth}
\centering
\includegraphics[width=8cm,height=7cm]{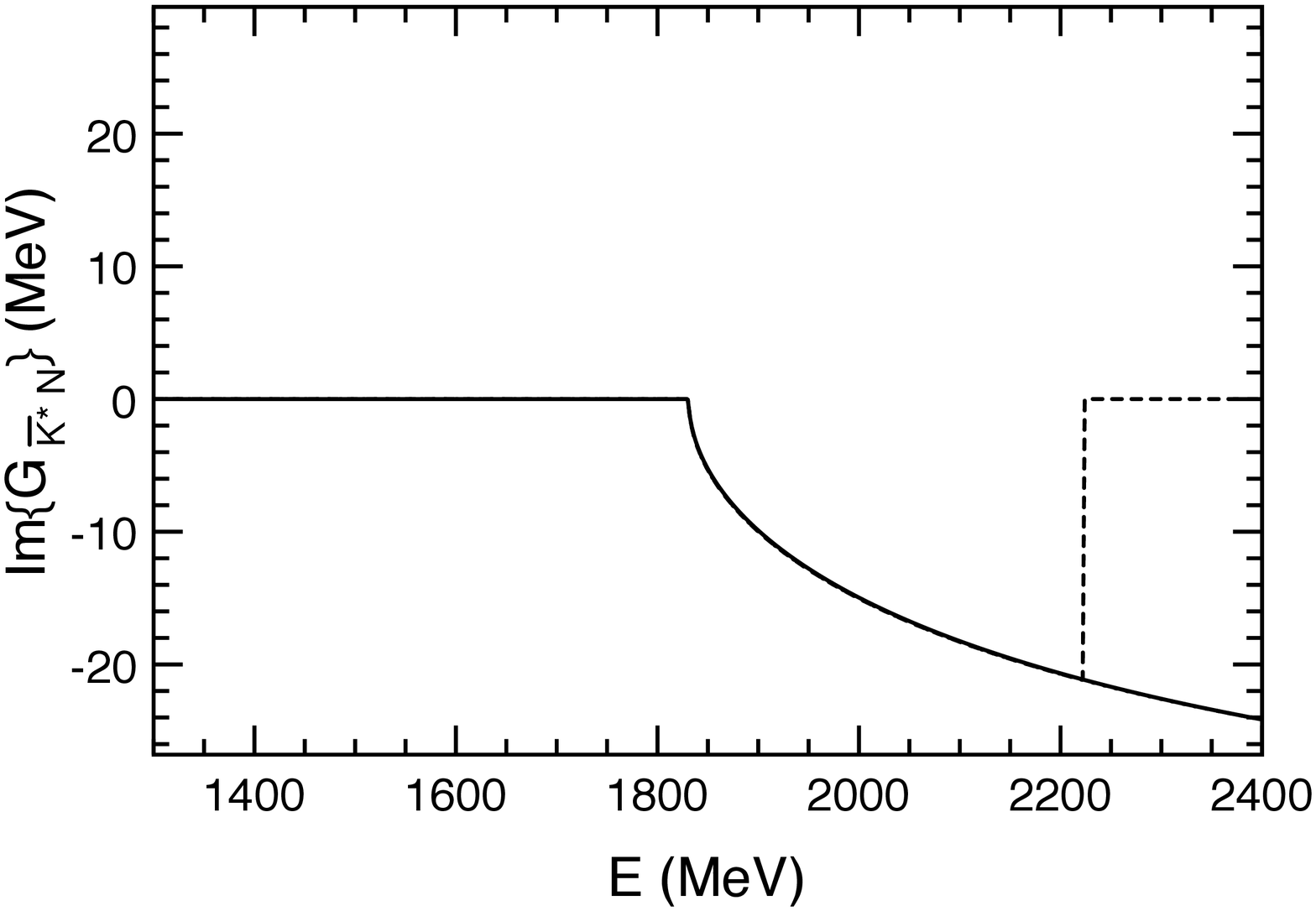}
\end{minipage}
\caption{Real (left panel) and imaginary  (right panel) parts of the loop calculated for the $\bar{K}^* N$ channel in the real plane.}\label{loop}
\end{figure}

\begin{figure}[h!]
\includegraphics[width=0.49\textwidth]{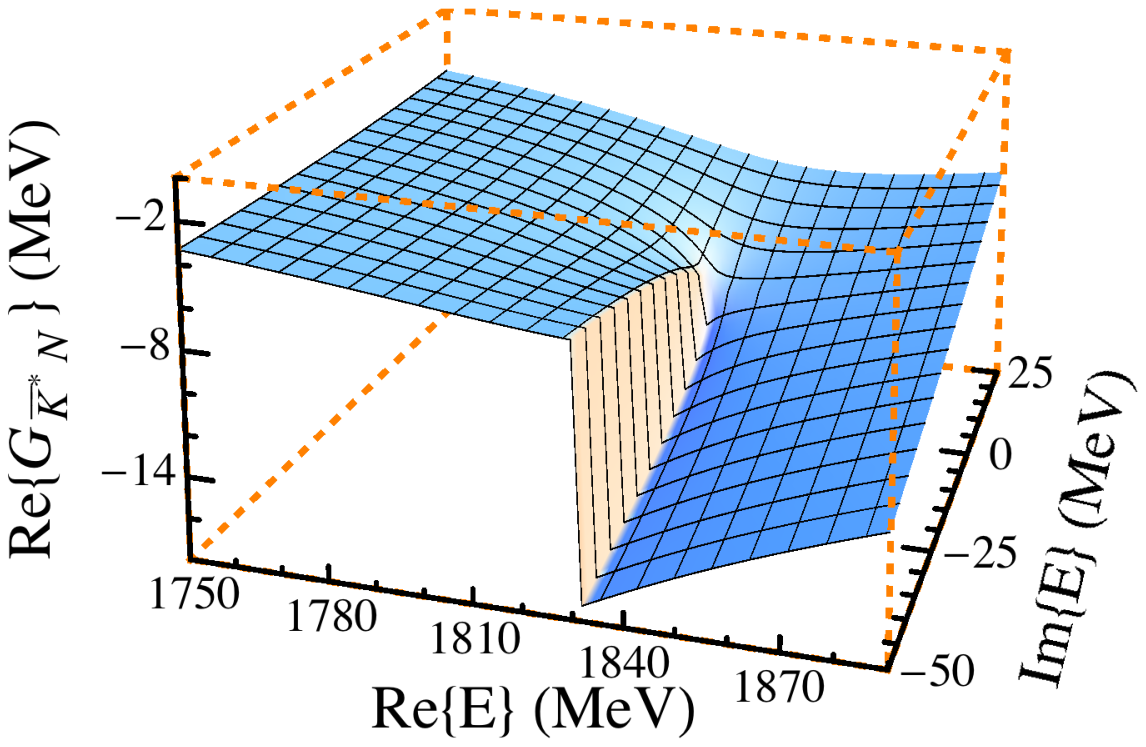}
\includegraphics[width=0.49\textwidth]{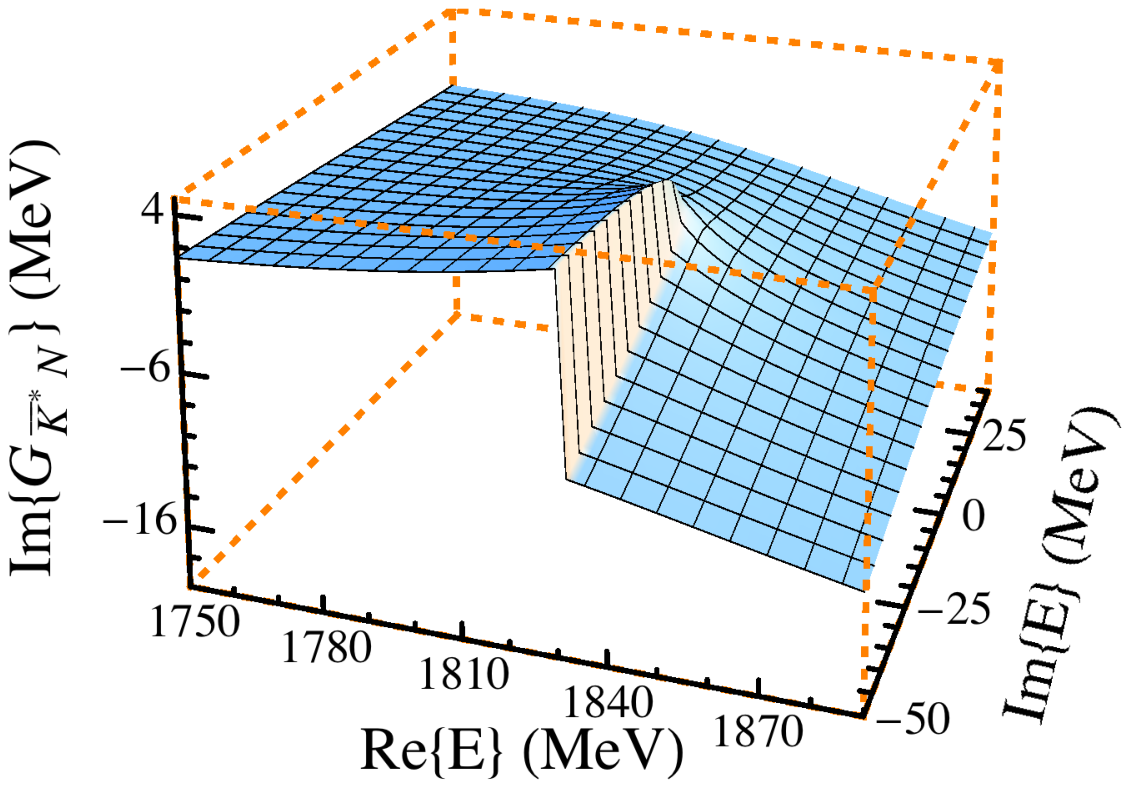}
\caption{Real (left panel) and imaginary  (right panel) parts of the loop (Eq.~(\ref{loopcutoff})) calculated for the $\bar{K}^* N$ channel in the second Riemann sheet. }\label{loop_complex}
\end{figure}

As an example, we show  the real part of the loop for the $\bar{K}^* N$ channel in the left panel of Fig.~\ref{loop}. The dotted line shows  the result of the calculation done by using the dimensional regularization method with a subtraction constant $a_{\bar{K}^* N} = -2$ and the dashed curve shows the loop calculated by using a cut-off of  630 MeV. This value of the cut-off is chosen such that the two schemes give the same result at the threshold.   It is interesting to see that the loop obtained within the dimensional regularization scheme changes the sign and becomes large and positive at lower energies. As discussed above, such a behavior of the loop is not what one would expect at the energies about 400-500 MeV below the threshold. We thus find that the use of the dimensional regularization scheme should be limited to  energies close to the threshold. Consequently, we must proceed with the cut-off method, which does not suffer this limitation. However, the loops calculated  with a cut-off  value of  630 MeV, which  gives the same result as the dimensional regularization at the threshold, possess  sharp structures (see the dashed lines in Fig.~\ref{loop} near 2.2 GeV) precisely due to the use of a sharp cut-off.  This would limit our calculations to the energies below  the one corresponding to the cut-off momentum. In order to solve this problem, we multiply the loops by a form factor
\begin{equation}
\mathcal{F}= e^{-\frac{\left(q^2 - q_{on}^2 \right)}{\Lambda^2}}, \label{ff}
\end{equation}
and extend the upper limit of the loop integral to infinity,
 which would correspond to taking the fact into account that the hadrons possess an extended spatial structure. In Eq.~(\ref{ff}), $q_{on}$ is the on-shell momentum in the center of mass system of the propagating meson-baryon pair and $\Lambda$ is the cut-off momentum.
The loop 
 \begin{equation}
 G = \int\limits_0^\infty \frac{d^3q}{\left(2 \pi \right)^3} \frac{1}{2 E_1 \left(\vec{q}, m\right)} \frac{2 M}{2 E_2 \left(\vec{q}, M\right)} \frac{\mathcal{F}}{E - E_1 \left(\vec{q}, m\right) - E_2 \left(\vec{q}, M\right)}, \label{loopcutoff}
 \end{equation} 
calculated in the non-relativistic approximation, and using a cut-off of 545 MeV and the form factor of Eq.~(\ref{ff}),  is shown  as the solid curve in 
Fig.~\ref{loop}.  In Eq.~(\ref{loopcutoff}), $E$ refers to the total energy of the meson-baryon system, which corresponds to the $E$ in Eq.~(\ref{pertE}). The loop shown in Fig.~\ref{loop} has been calculated using a cut-off 545 MeV because, as we shall show later, this value gives very similar amplitudes as those obtained by the dimensional regularization method.  It can be seen in Fig.~\ref{loop}  that the real part of the loop calculated in this way does not change sign even when going up to very low energies. Also, the imaginary part of the loop shows the expected characteristics. We further show  that the behavior of the loops, calculated using Eq.~(\ref{loopcutoff}), in the complex plane is as expected. For this we depict the loop in the second Riemann sheet for the $\bar{K}^* N$ channel  in Fig.~\ref{loop_complex}. 

To end with the discussion of the loops, we would like to mention that in  case of systems involving vector mesons, which can possess large widths sometimes, like the $\rho$ and the $K^*$-mesons, we calculate the  loop given by Eq.~(\ref{loopcutoff}), by convoluting them over the mass range of these mesons as discussed in Ref.~\cite{eulogiovb}. 

\begin{figure}[h]
\begin{minipage}[b]{0.49\linewidth}
\centering
\includegraphics[width=8cm,height=7cm]{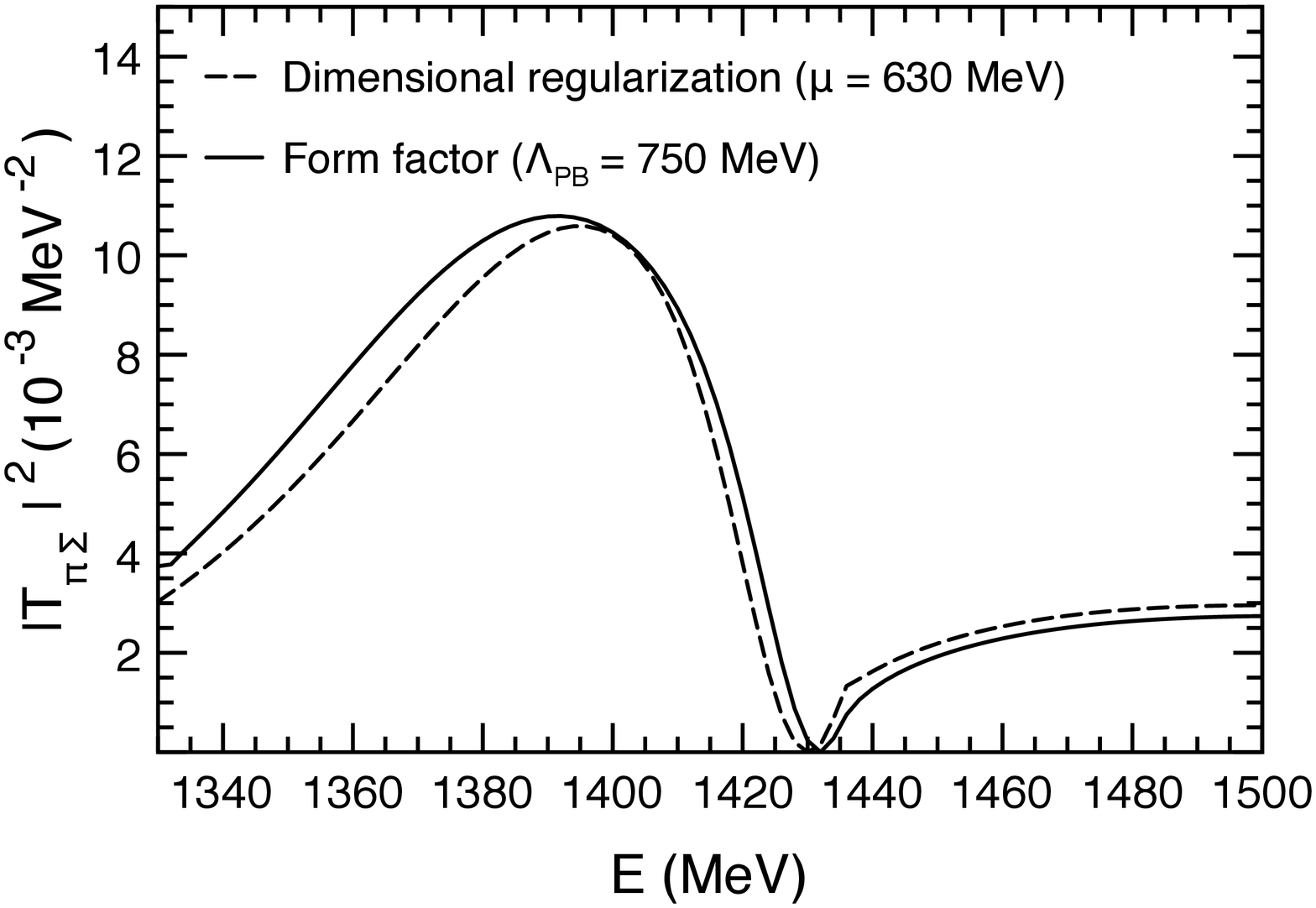}
\end{minipage}
\begin{minipage}[b]{0.49\linewidth}
\centering
\includegraphics[width=8cm,height=7cm]{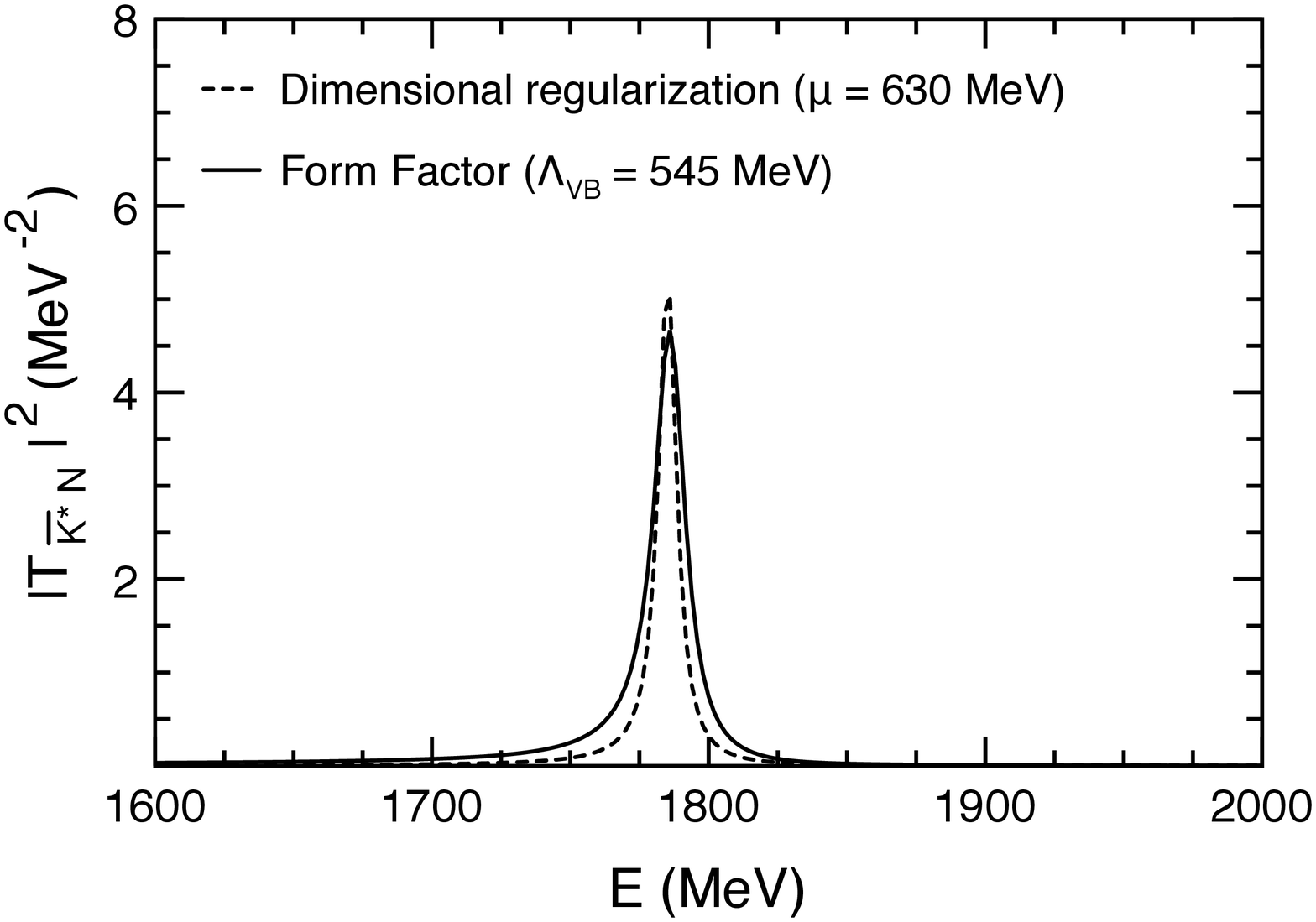}
\end{minipage}
\caption{ Squared amplitude of the $\pi \Sigma$ (left panel) and $\bar{K}^* N$ (right panel) channels obtained by considering uncoupled PB-VB systems. The dashed and solid lines are the results of the calculation done with the loops obtained with the dimensional regularization and cut-off  method (Eq.~(\ref{loopcutoff})), respectively.}\label{ampchk}
\end{figure}

Further, to test the reliability of our method to calculate the loop, we obtain the amplitudes for the uncoupled PB and VB systems with 
the loop given by Eq.~(\ref{loopcutoff}). We found that in order to reproduce the results of previous uncoupled studies of PB and VB  systems, which agree with the available experimental data, we need to use a cut-off $\Lambda_{VB} =$ 545 MeV for the VB  and  $\Lambda_{PB} =$ 750 MeV for PB systems. Using these two parameters, we have found that we can reproduce all the previous results of Refs.\cite{eulogiopb, eulogiovb} reasonably well. As an example, in Fig.~\ref{ampchk}, we show a comparison of the  $\pi \Sigma$ and $\bar{K}^* N$ amplitudes  obtained by solving the coupled channel Bethe-Salpeter equation with the loops calculated by using the dimensional regularization method and by using the cut-off method involving a form factor (Eq.~(\ref{loopcutoff})). As can be seen in Fig.~\ref{ampchk}, the two results are satisfactorily similar. In the following, we will also discuss the poles obtained in the uncoupled PB-VB systems, along with their coupling to the related channels, in the calculations done with the loops obtained by using Eq.~(\ref{loopcutoff}).
 
\section{Results and discussions}
\subsection{Coupled PB-VB systems in Isospin 0}
\begin{figure}[ht!]
\begin{tabular}{cc}
\includegraphics[width=0.4\linewidth,height=4.5cm]{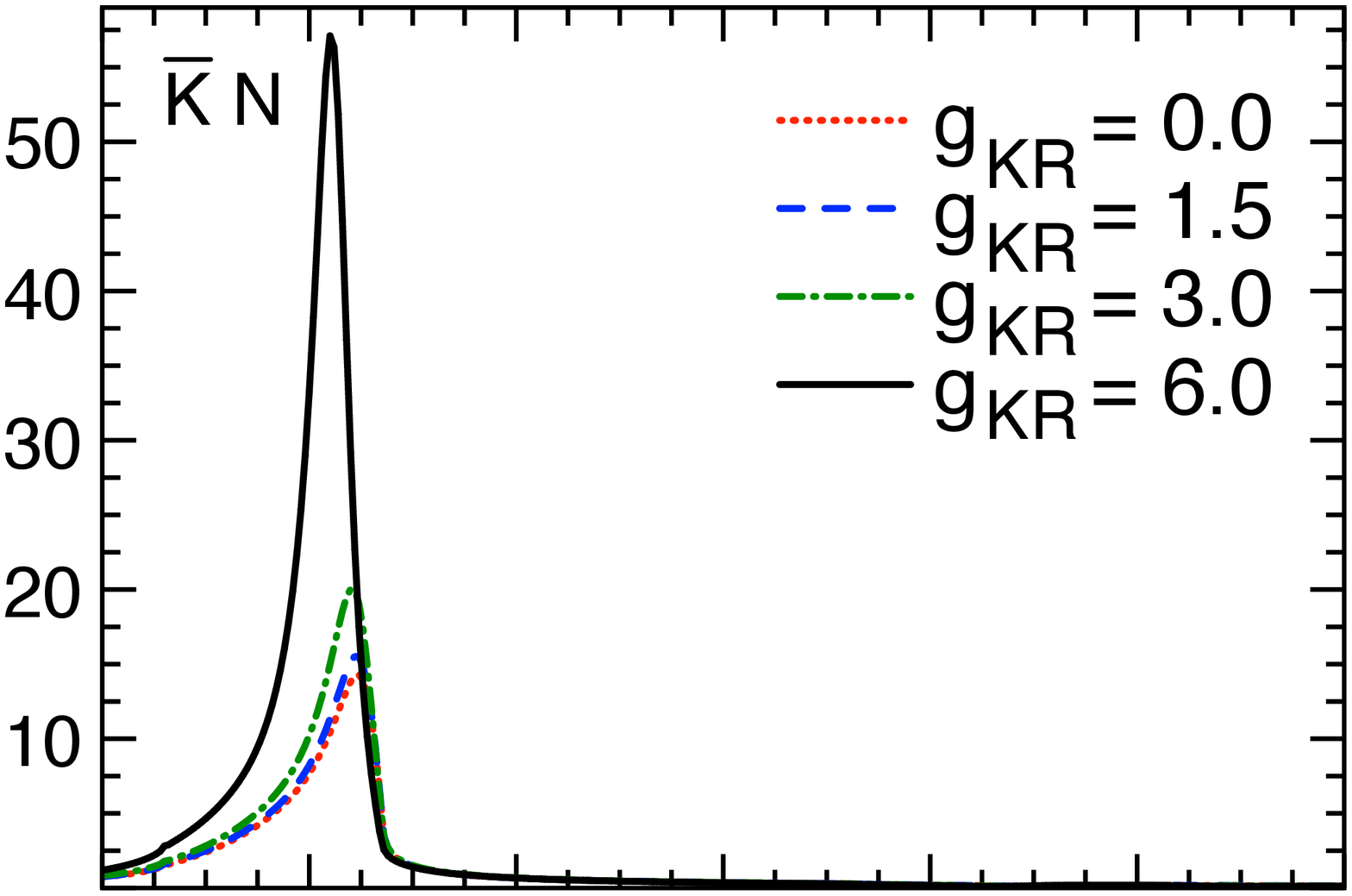}&\includegraphics[width=0.4\linewidth,height=4.5cm]{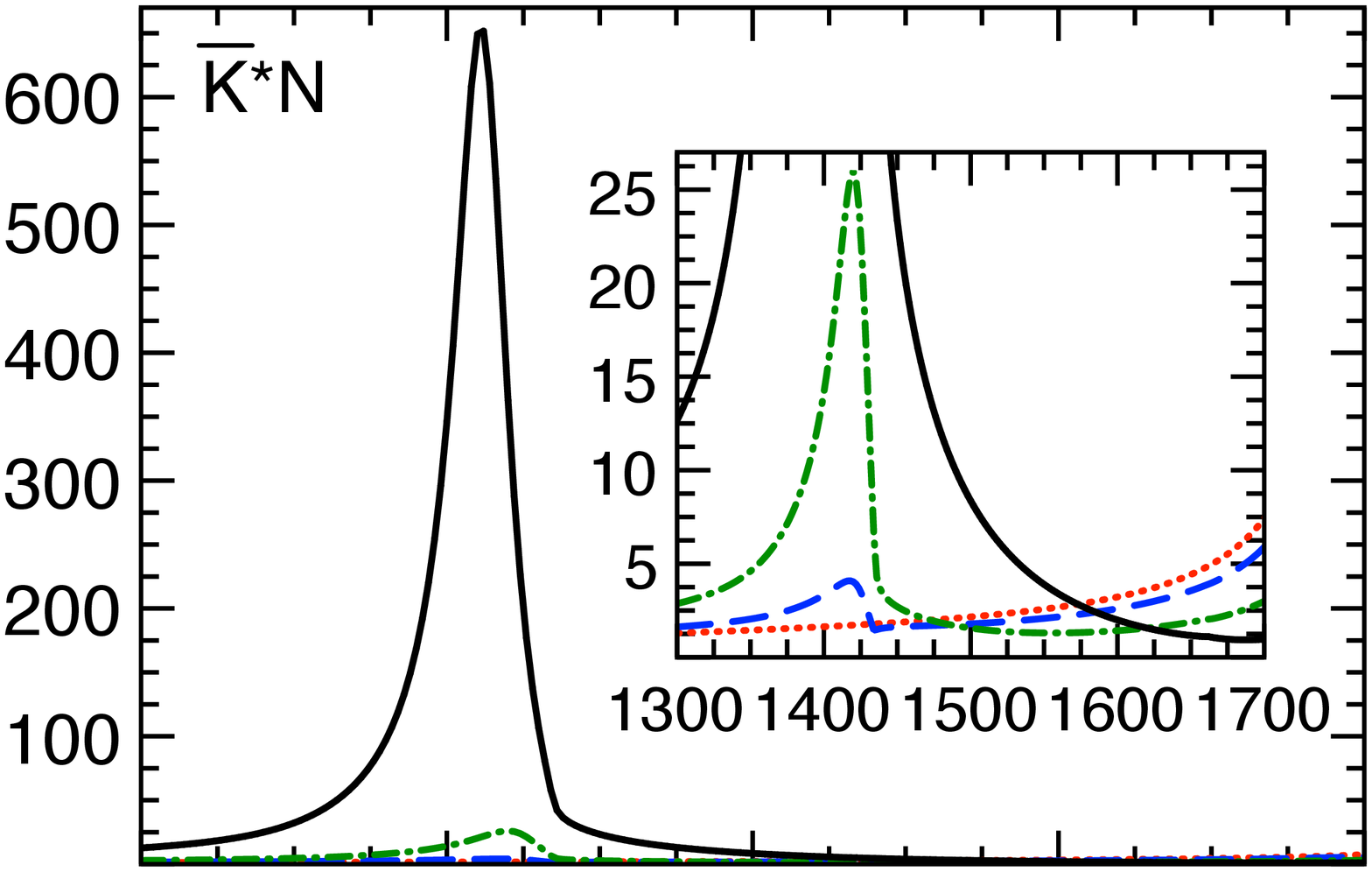}\\[-0.93cm]
\includegraphics[width=0.4\linewidth,height=4.5cm]{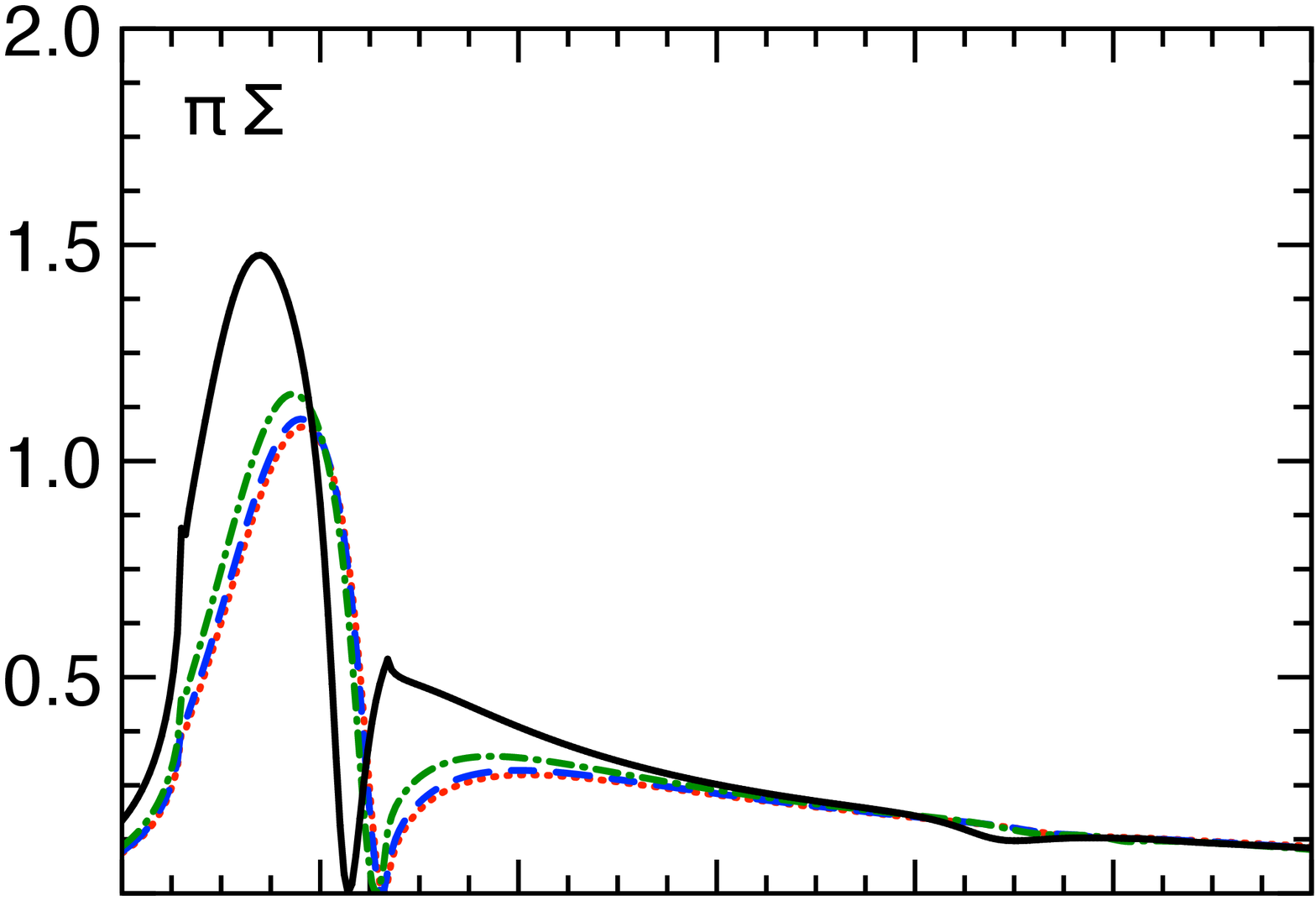}&\hspace{0.12cm}\includegraphics[width=0.4\linewidth,height=4.5cm]{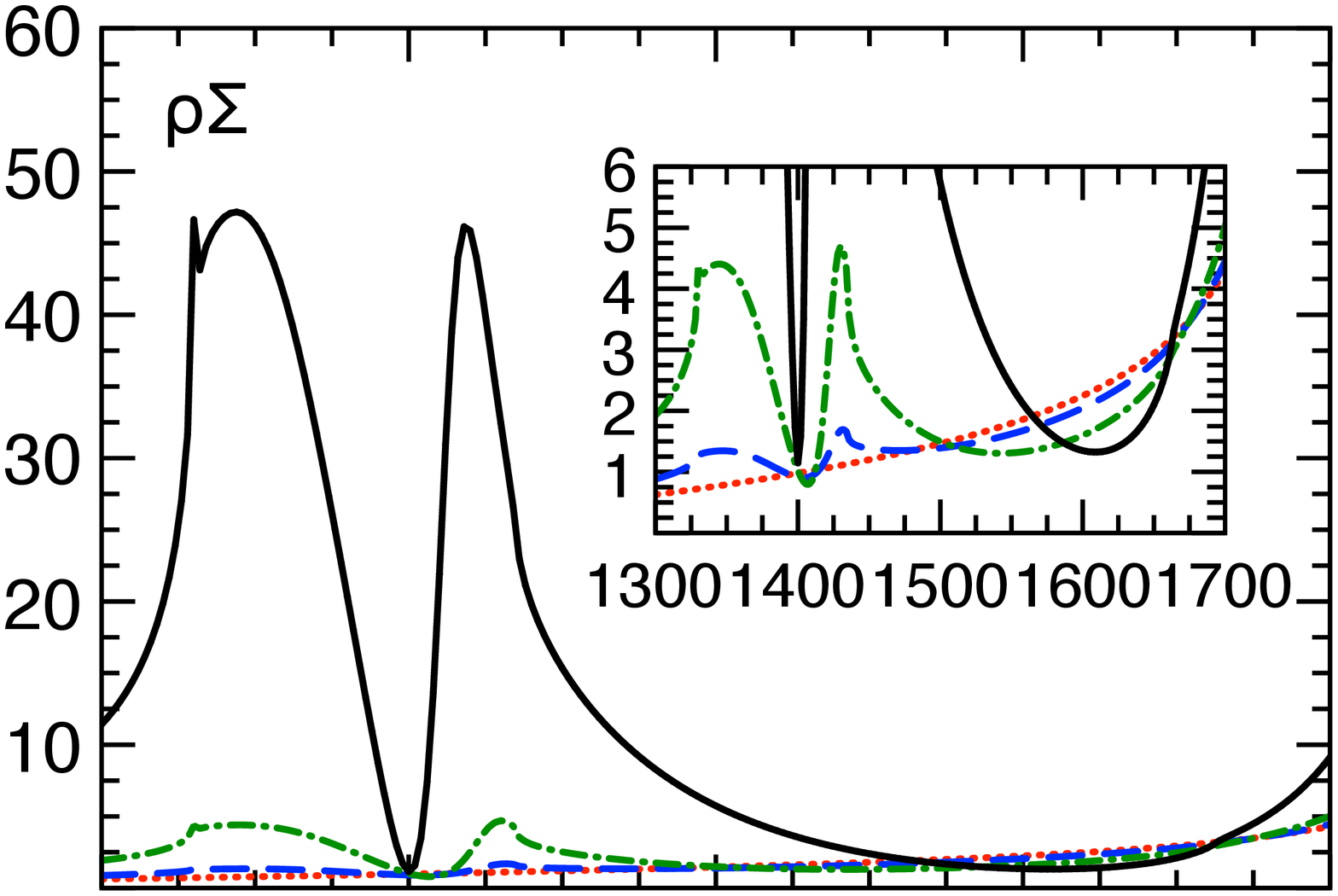}\\[-0.93cm]
\hspace{-1.1cm}
\includegraphics[width=0.455\linewidth,height=4.5cm]{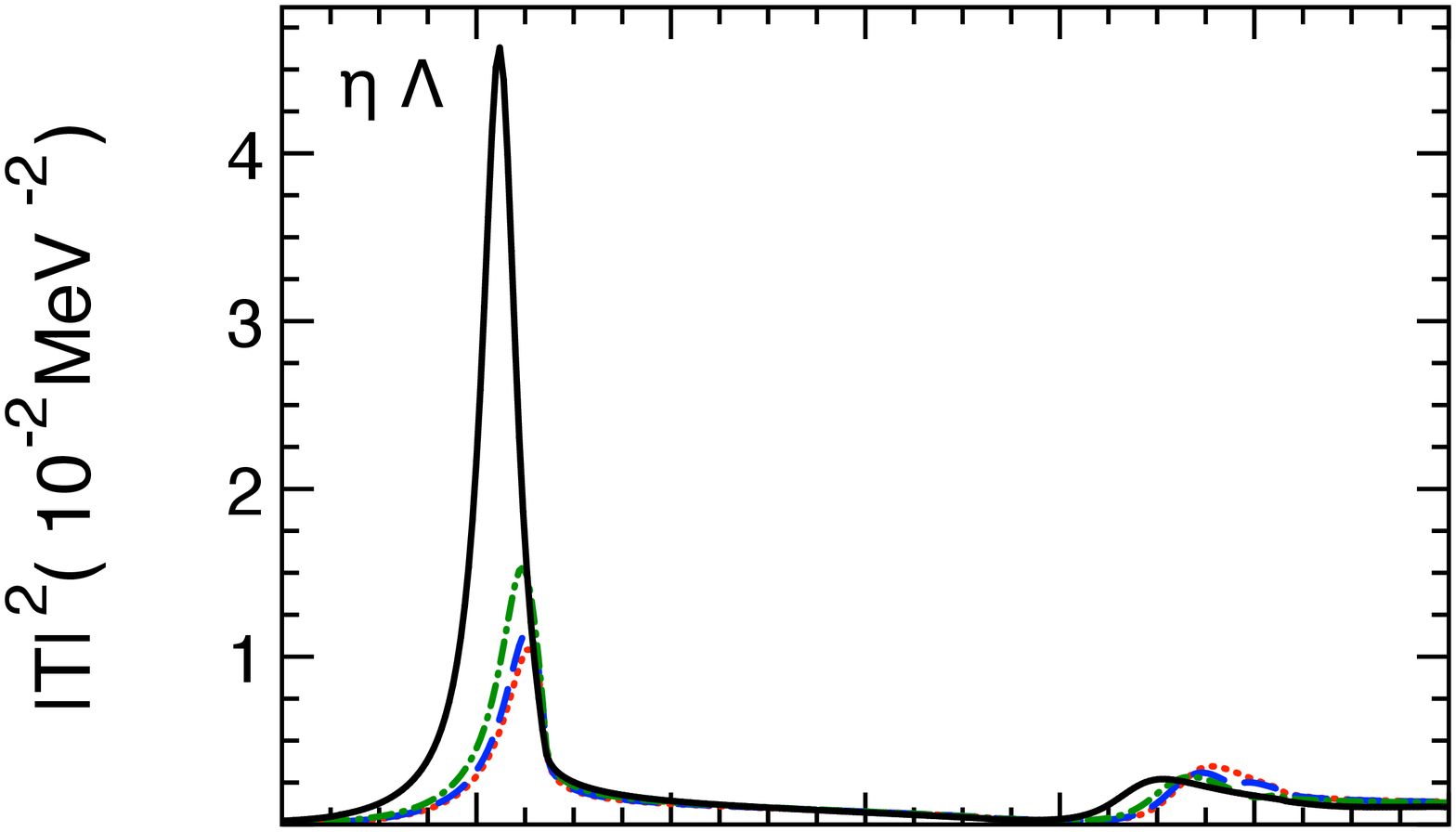}&\includegraphics[width=0.4\linewidth,height=4.5cm]{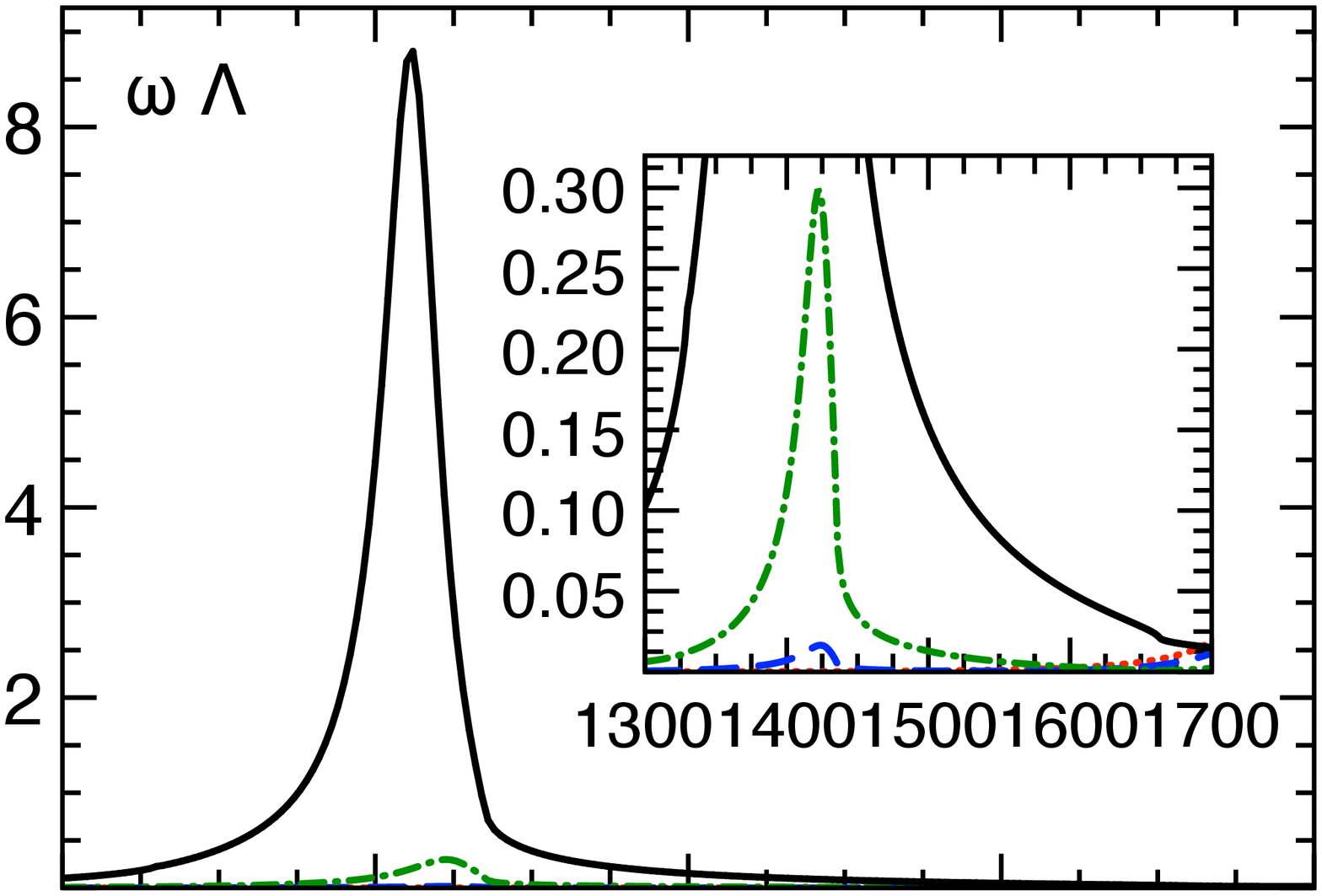}\\[-0.93cm]
\includegraphics[width=0.397\linewidth,height=5.1cm]{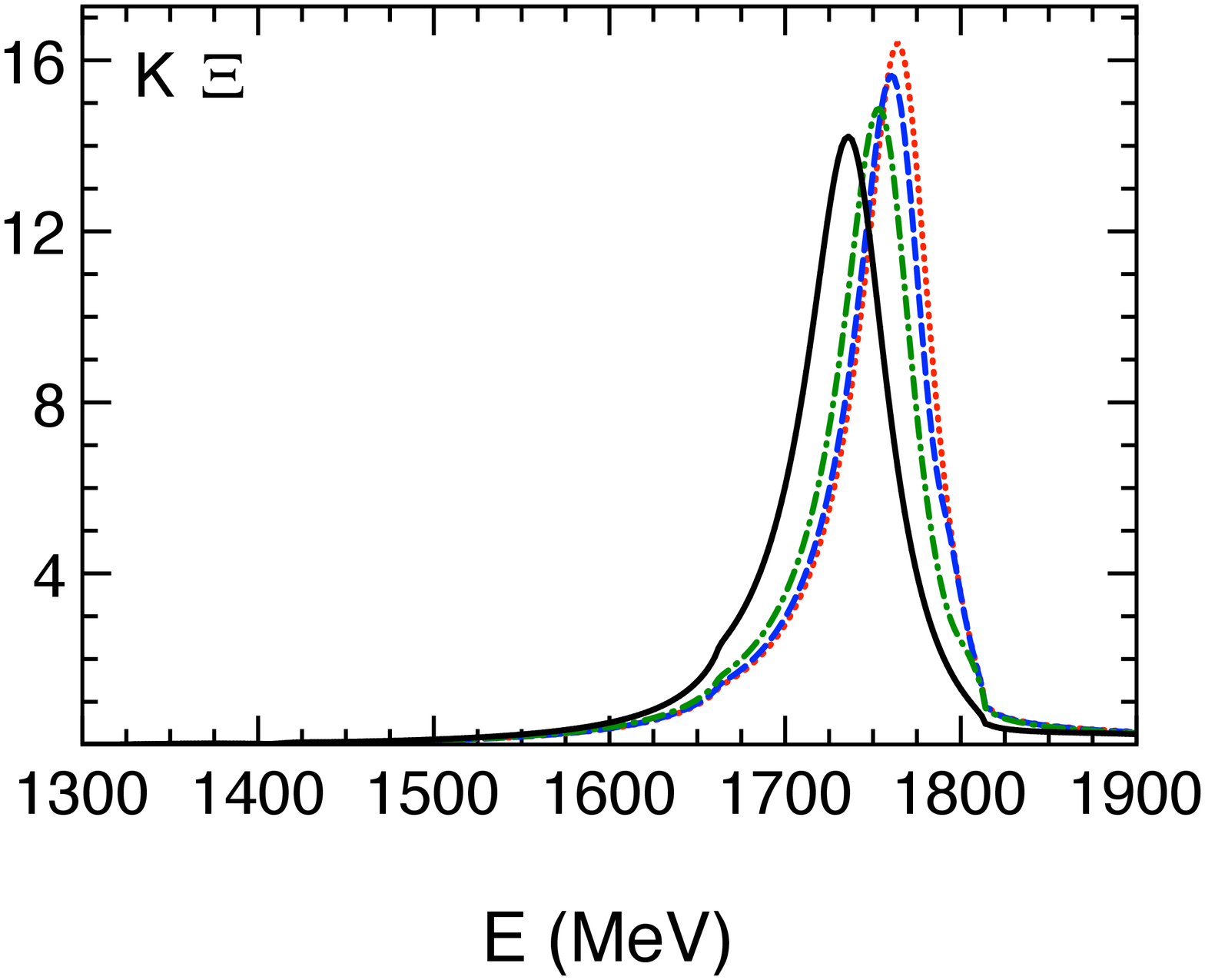}&\hspace{0.12cm}\includegraphics[width=0.4\linewidth,height=5.1cm]{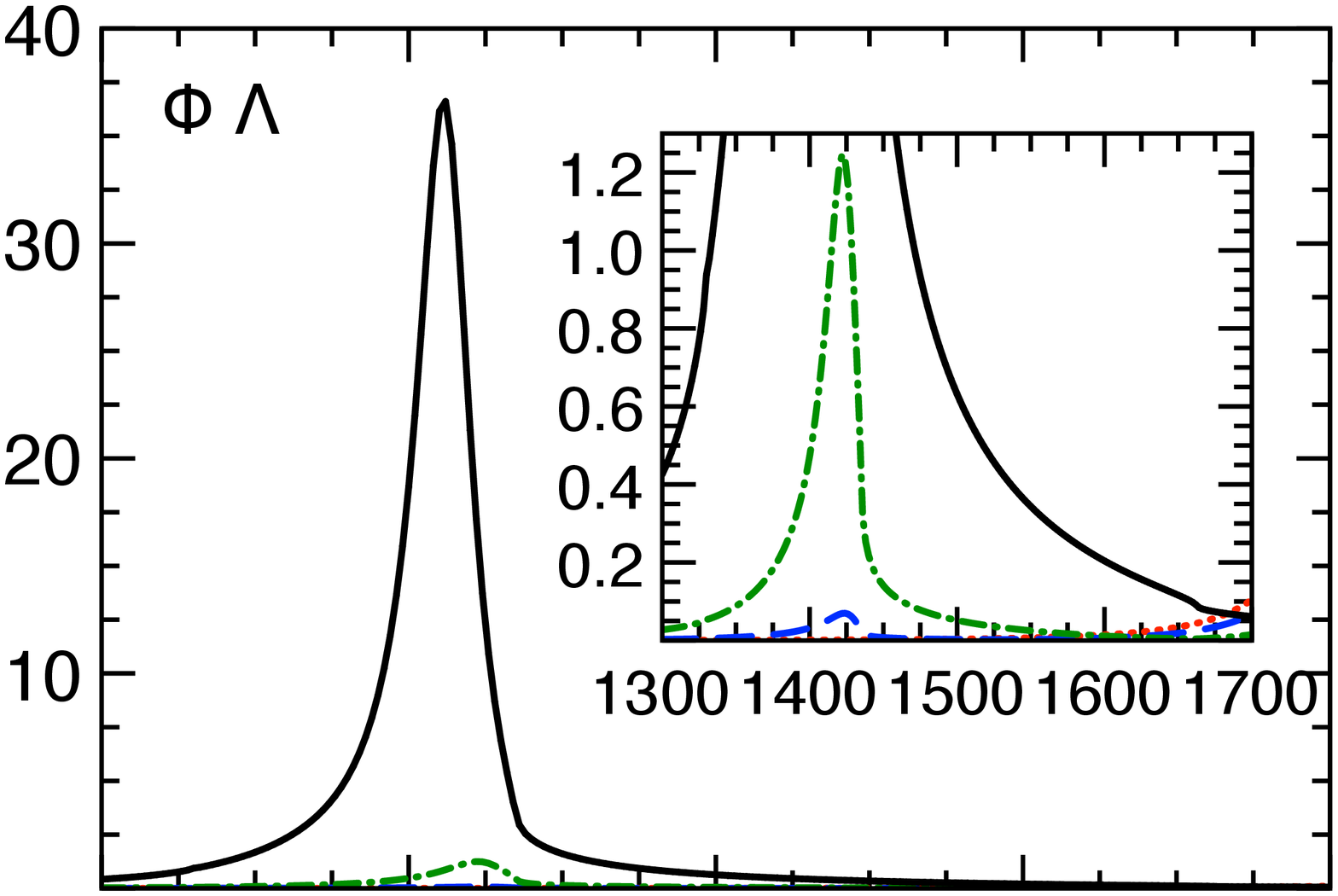}\\[-1.75cm]
&\includegraphics[width=0.4\linewidth,height=5.1cm]{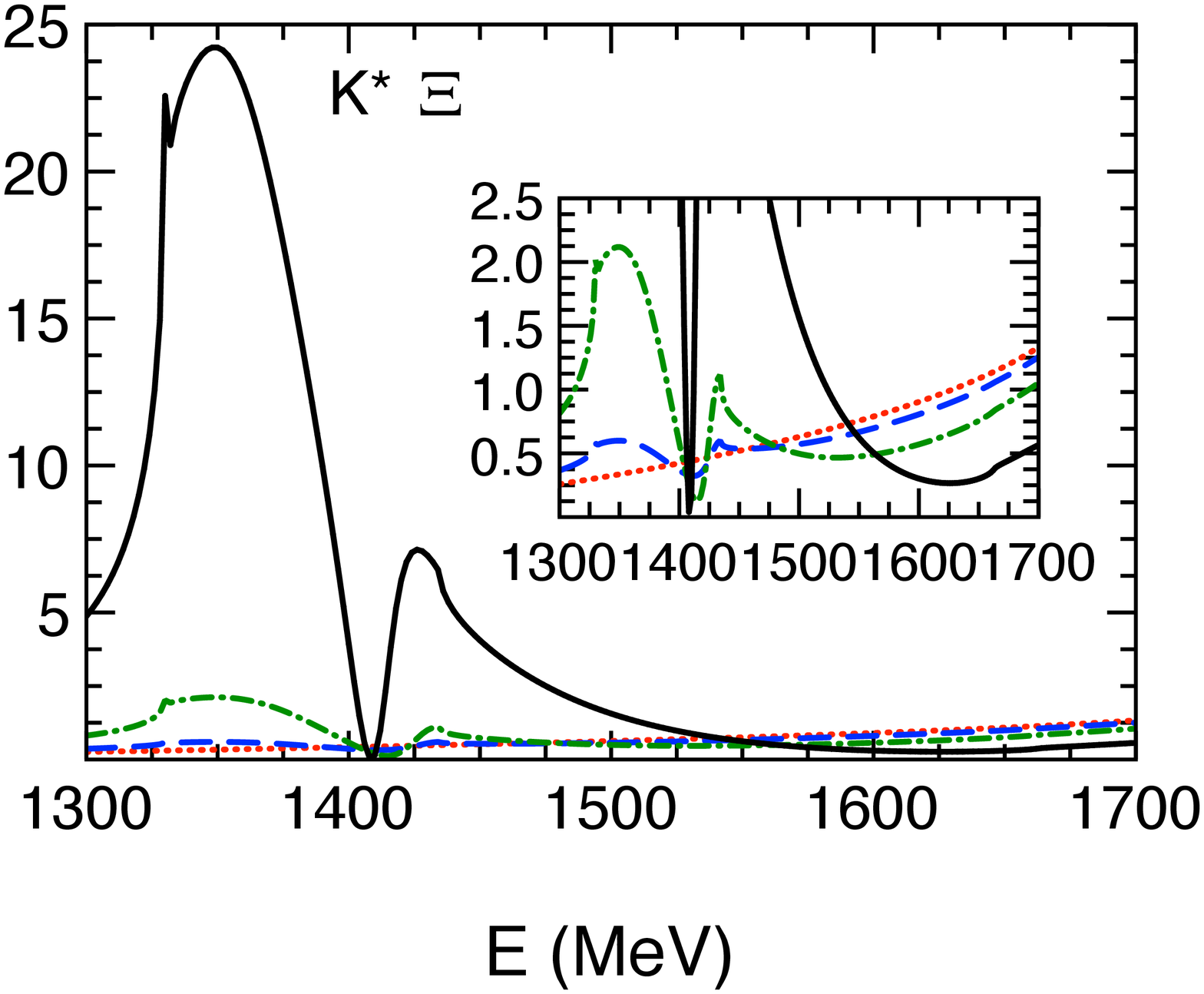}
\end{tabular}
\vspace{-0.5cm}
\caption{Isospin 0 amplitudes of the PB and VB systems for the energy $\leq$ 1900 MeV. The purpose of the inset figures is to show the structure of the amplitudes hidden in the corresponding large figures due to their smaller magnitudes.} \label{fig_iso0}
\end{figure}

\begin{figure}[ht!]
\centering
\begin{tabular}{cc}
\includegraphics[width=0.4\linewidth,height=4.4cm]{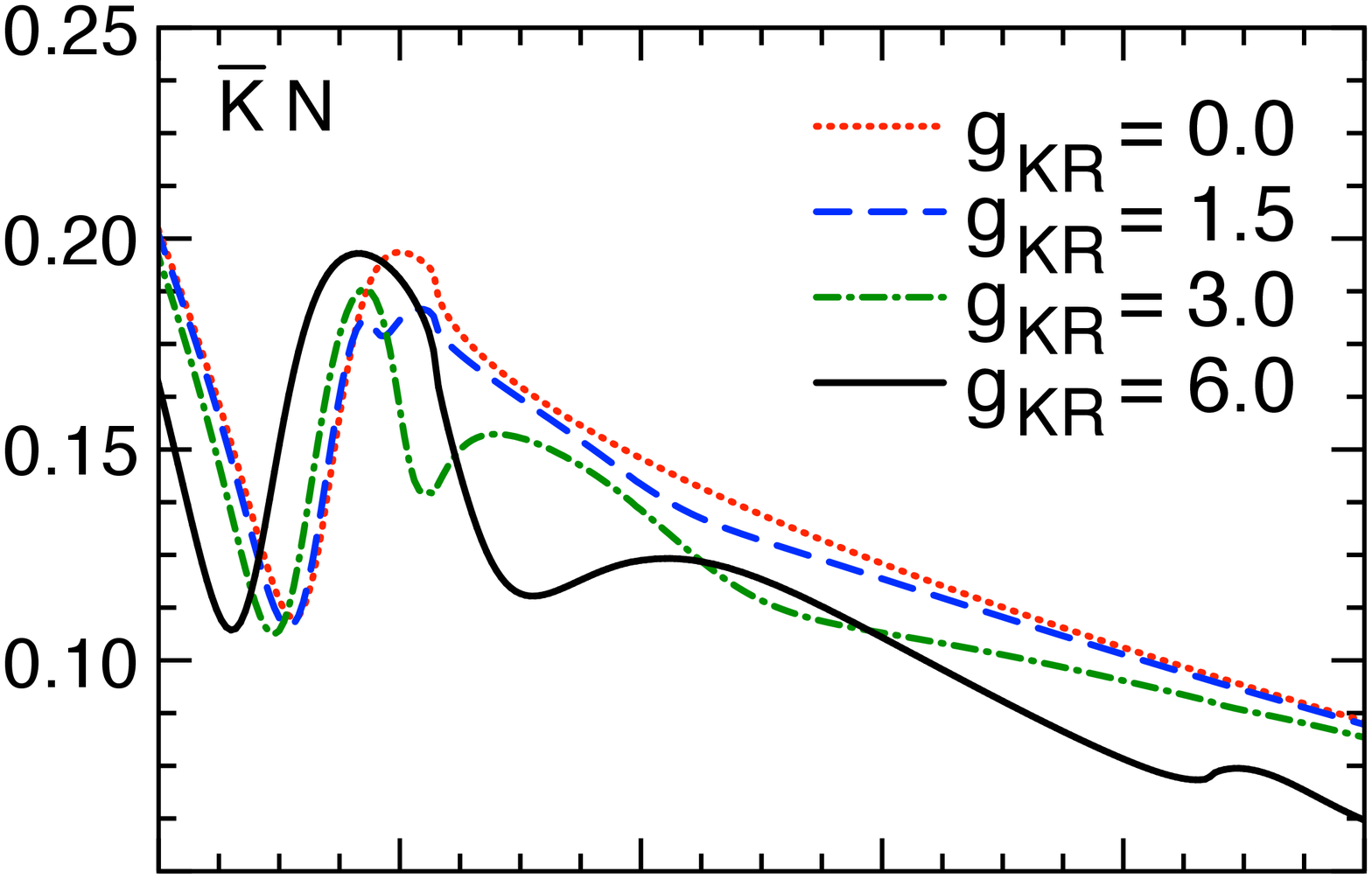}&\includegraphics[width=0.4\linewidth,height=4.4cm]{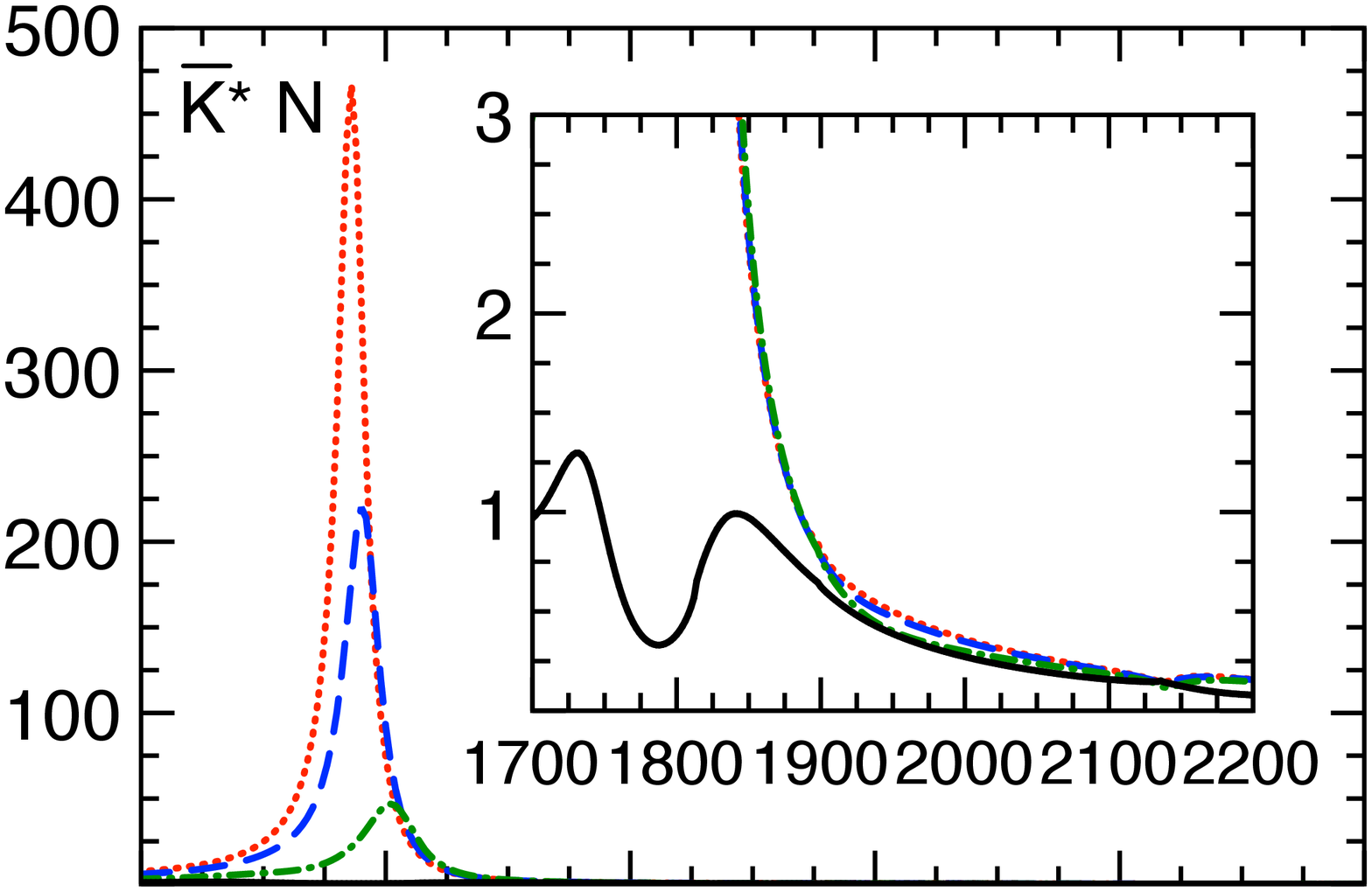}\\[-0.9cm]
\includegraphics[width=0.4\linewidth,height=4.4cm]{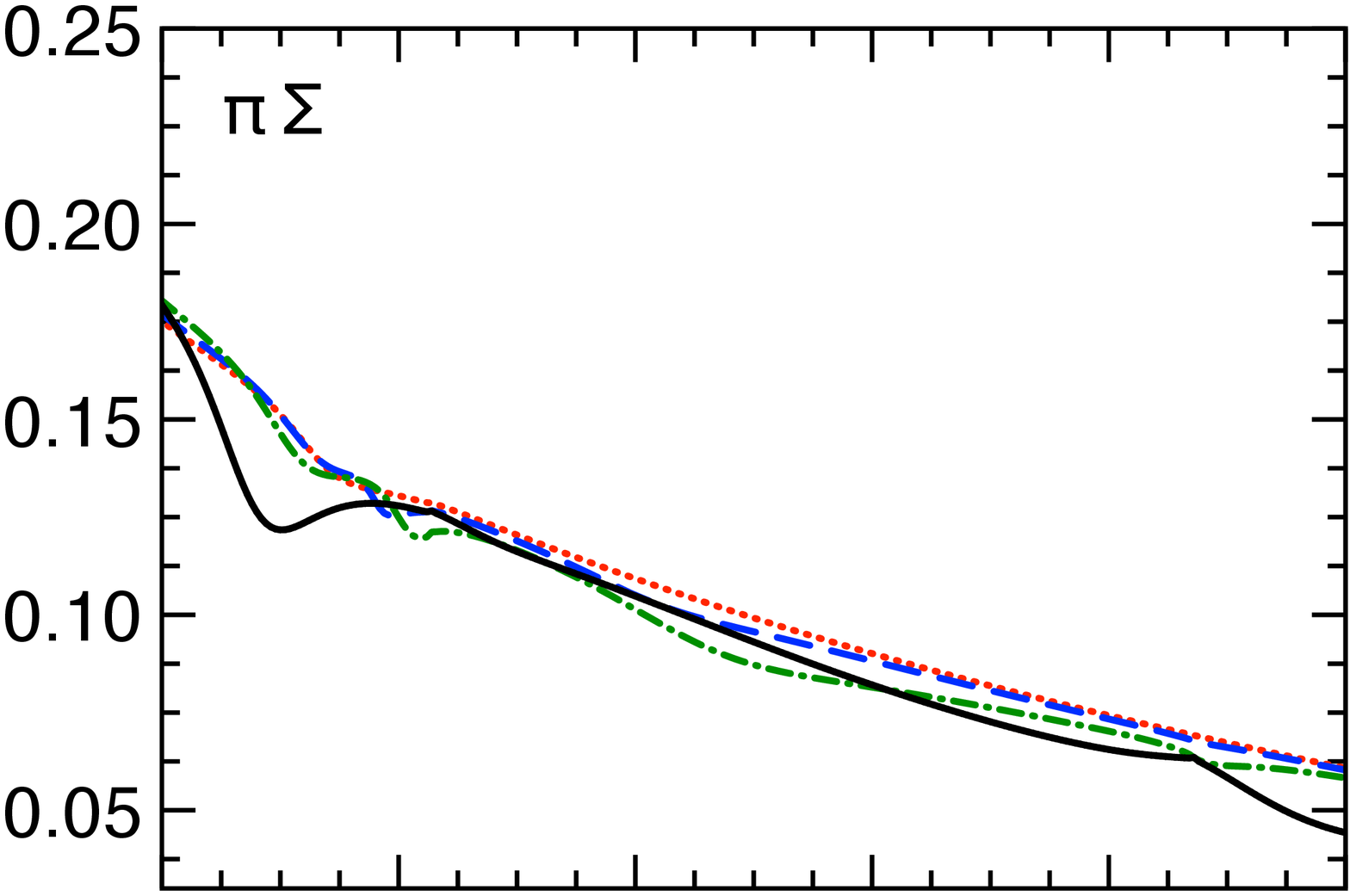}&\includegraphics[width=0.4\linewidth,height=4.4cm]{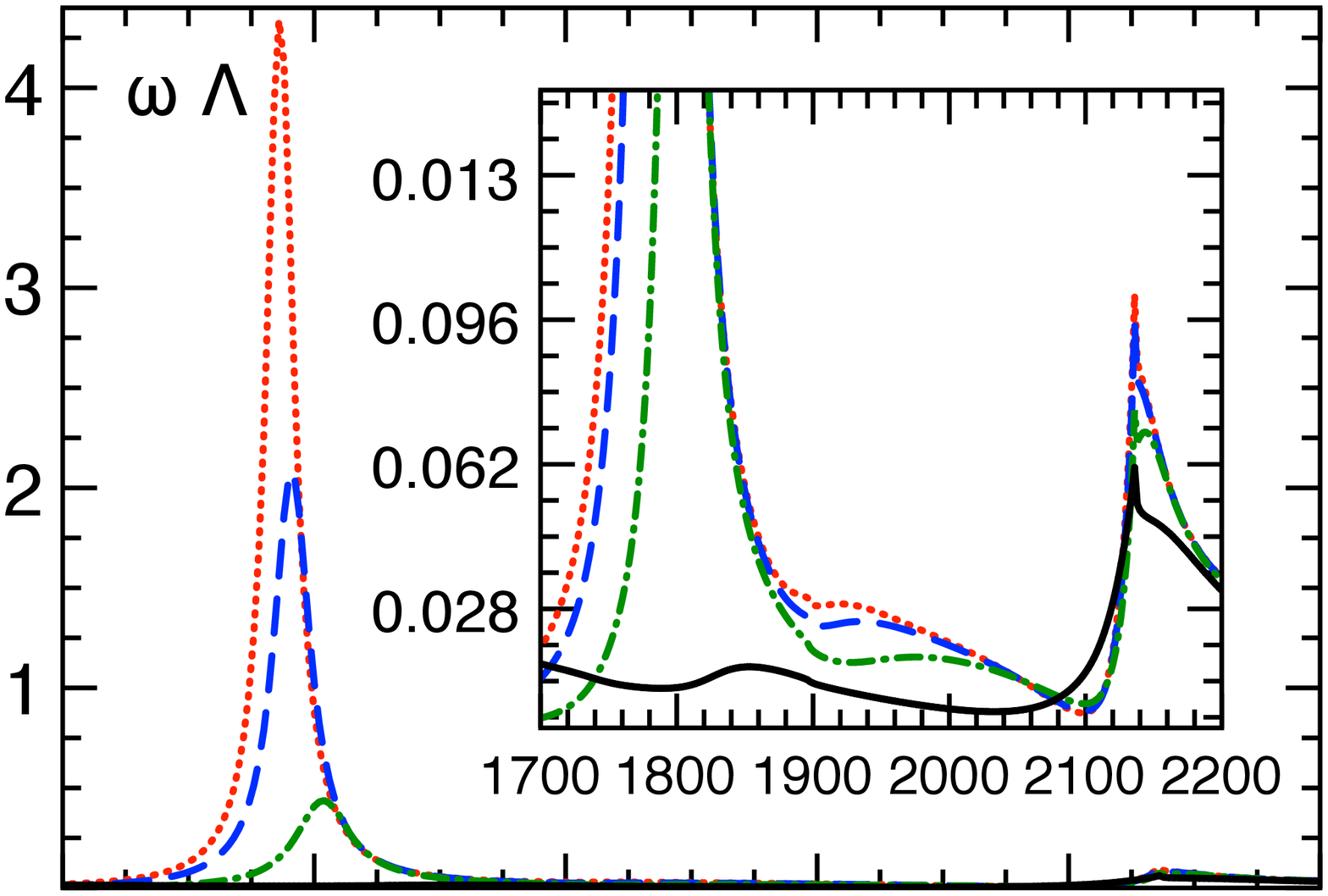}\\[-0.9cm]
\hspace{-0.96cm}\includegraphics[width=0.458\linewidth,height=4.4cm]{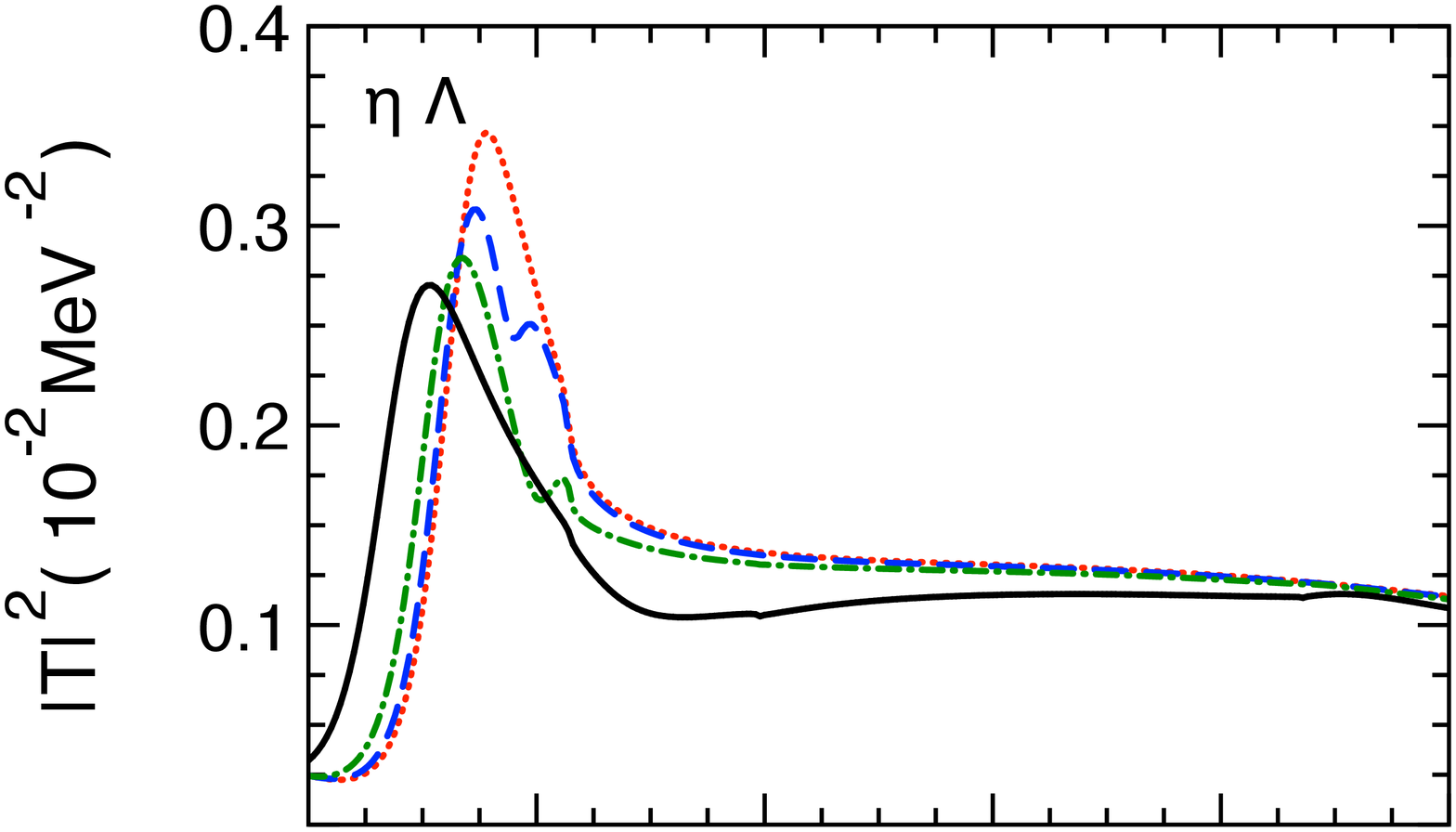}&\includegraphics[width=0.4\linewidth,height=4.4cm]{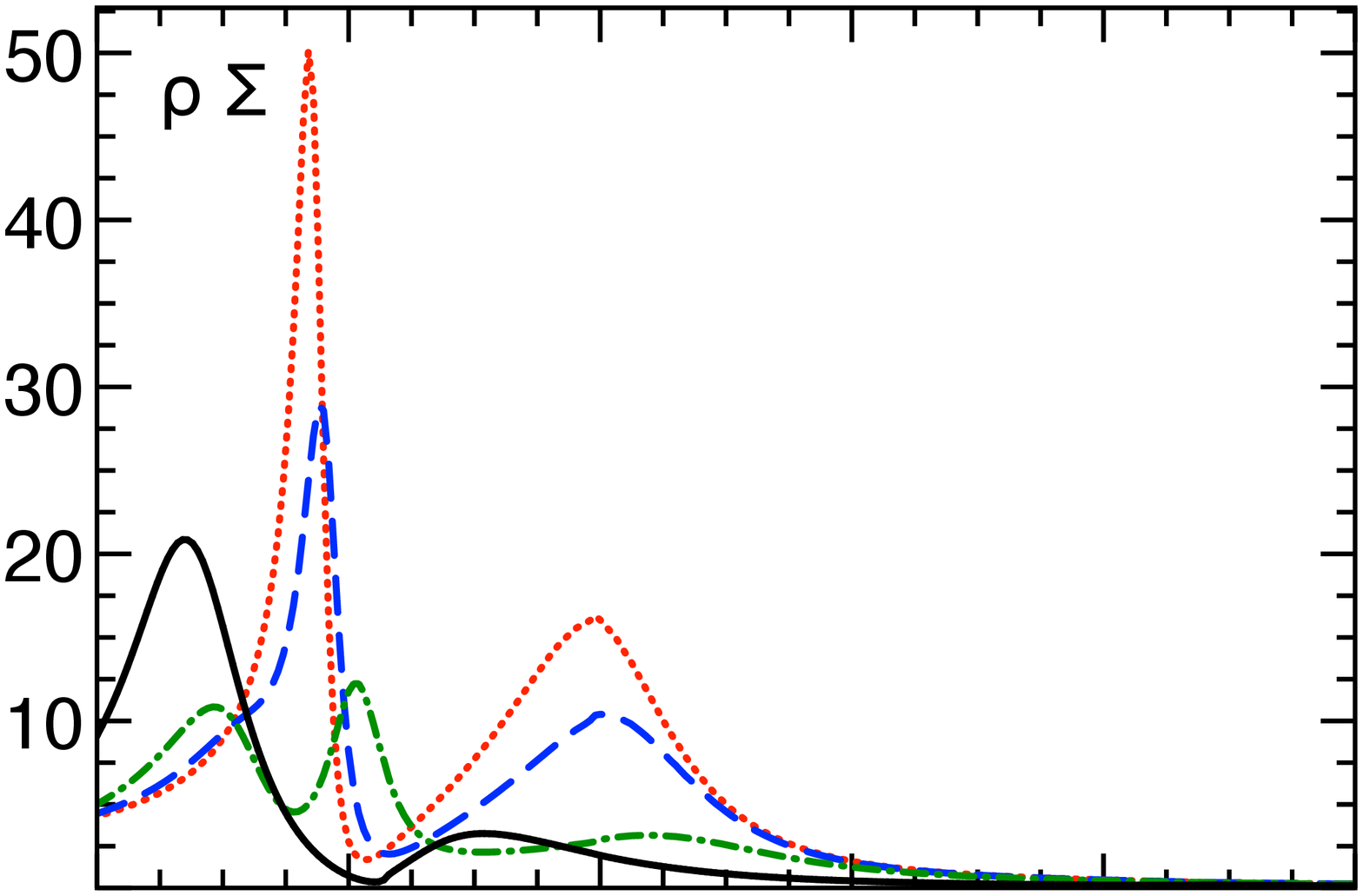}\\[-0.9cm]
\hspace{0.01cm}\includegraphics[width=0.40\linewidth,height=5.5cm]{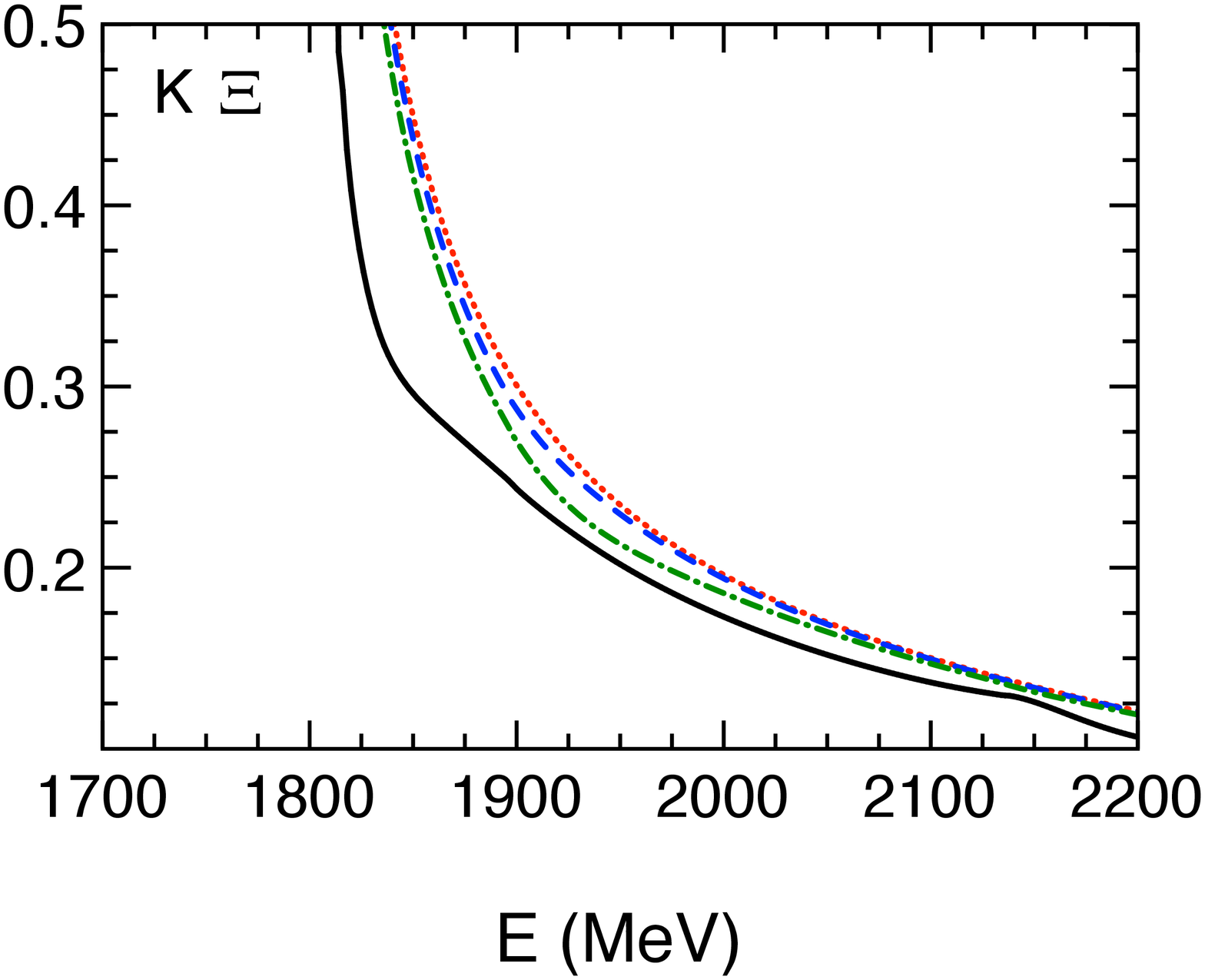}&\includegraphics[width=0.4\linewidth,height=5.5cm]{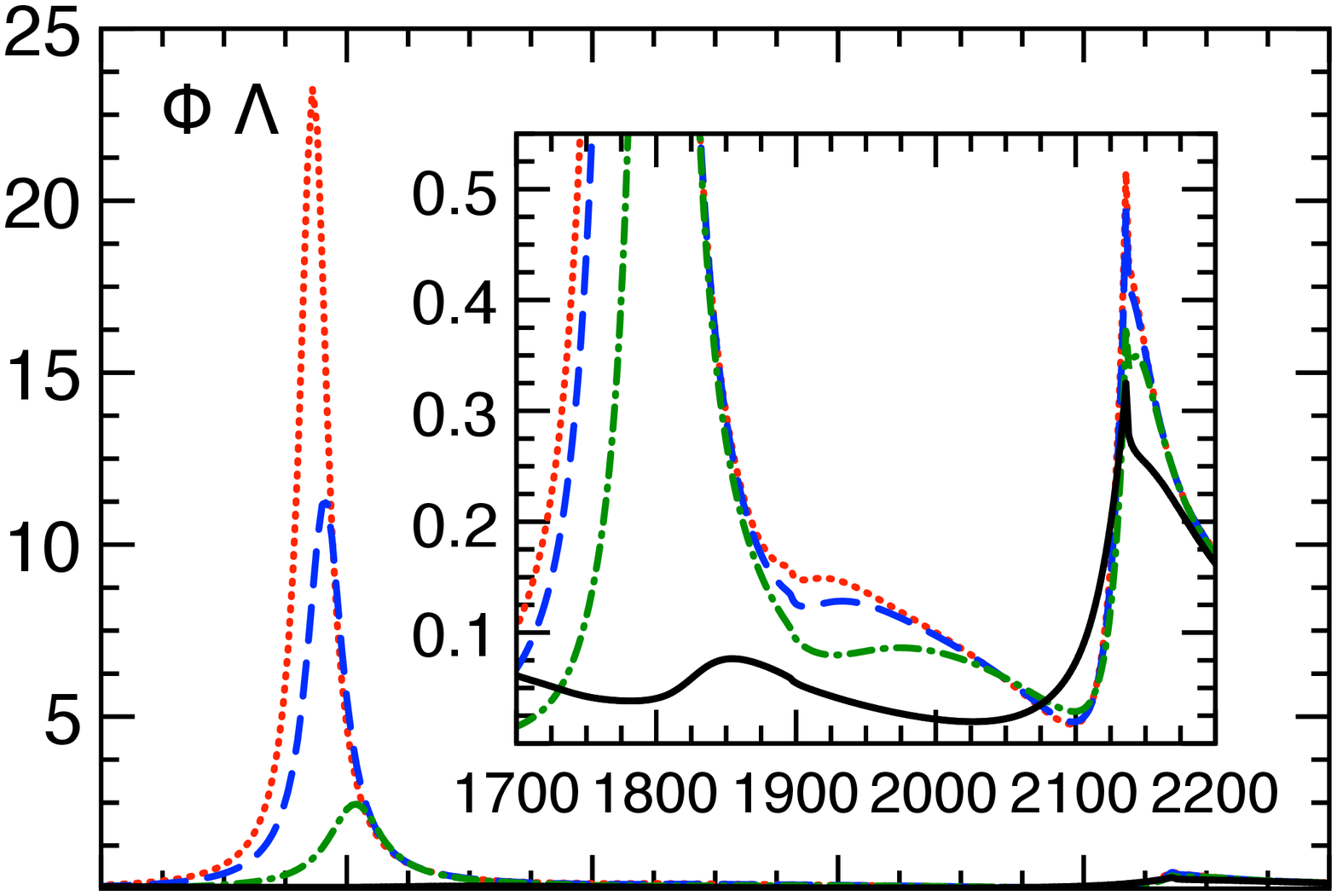}\\[-1.9cm]
&\includegraphics[width=0.405\linewidth,height=5.5cm]{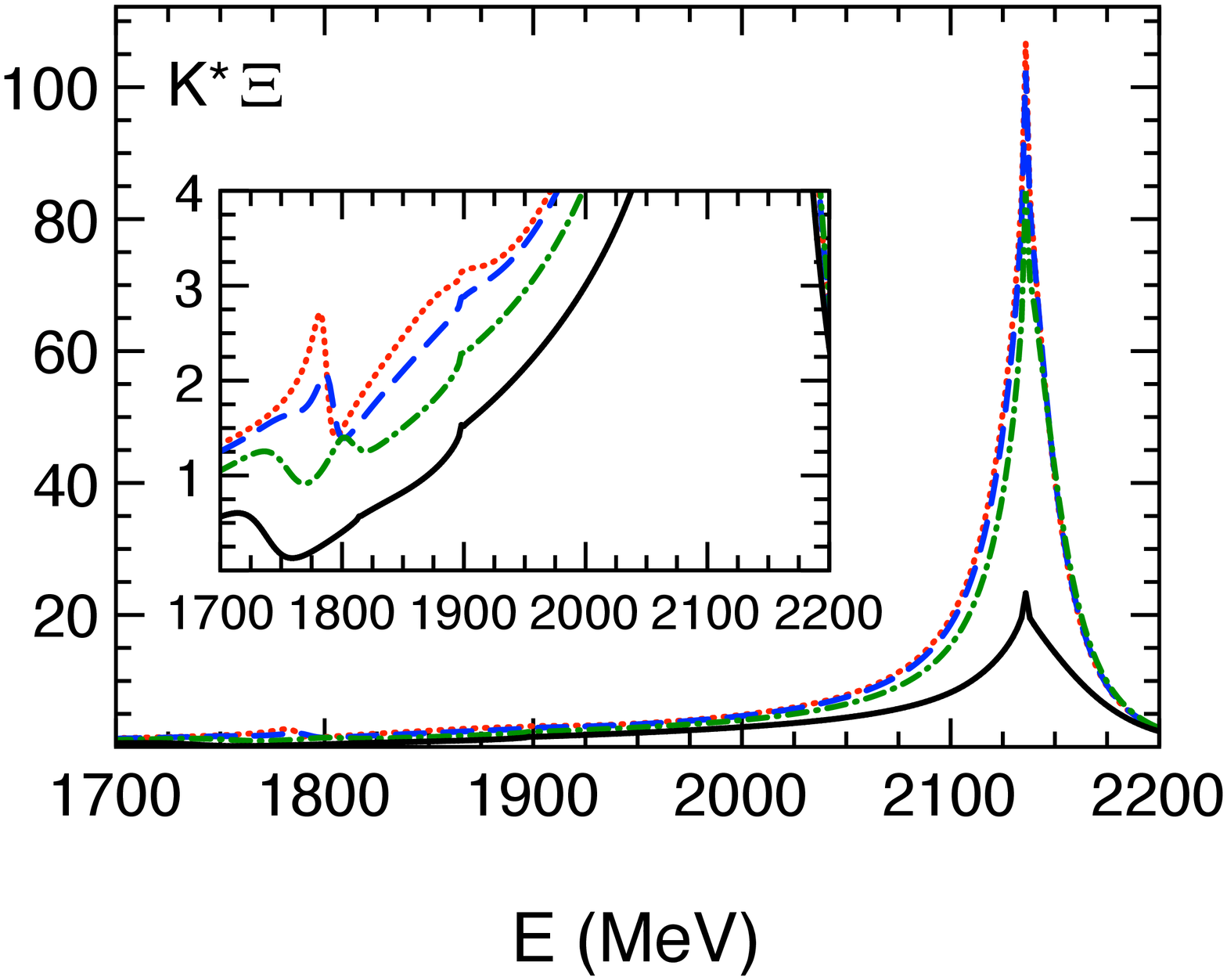}
\end{tabular}
\caption{Isospin 0 amplitudes of the PB and VB systems for the energy region 1700-2200 MeV. The inset figures here have the same objective as those in  Fig.~\ref{fig_iso0}.} \label{fig_iso0_II}
\end{figure}

Let us first study the total isospin zero systems, in which case we have nine (four PB and five VB) coupled channels:  $\bar{K} N$, $\pi \Sigma$, $\eta \Lambda$, $K \Xi$, $\bar{K^*}N$, $\omega \Lambda$, $\rho \Sigma$, $\phi \Lambda$ and $K^*\Xi$.  We first obtain the amplitudes keeping the coupling between the PB and VB channels, $g_{KR}$, as zero, i.e., treating them as uncoupled systems, and by calculating the loops (Eq.~(\ref{loopcutoff})) using  the cut-offs  $\Lambda_{VB}$ = 545 MeV for the VB loops and $\Lambda_{PB}$ = 750 MeV for the PB case. The amplitudes obtained on the real axis in this case are shown as the dotted curves in Figs.~\ref{fig_iso0}, \ref{fig_iso0_II}. The corresponding poles are listed in the second column of  Tables~\ref{iso0_1}-\ref{iso0_6}. There are six poles found in this case: two related to the $\Lambda (1405)$, one to  the  $\Lambda (1670)$ and the rest three are spin-degenerate in nature due to the spin-independent form of the vector-baryon interaction. Together,  Figs.~\ref{fig_iso0}, \ref{fig_iso0_II}
and  Tables~\ref{iso0_1}-\ref{iso0_6} show that the results obtained with $g_{KR} = 0$ are in good agreement with the ones reported in Refs.~\cite{eulogiopb, eulogiovb}.

Next, we change the coupling between the PB and VB channels, $g_{KR}$, from zero
to its maximum value which is 6 and show how the amplitudes and poles found in the uncoupled PB-VB systems ($g_{KR} =$ 0) change their behavior due to the 
coupling between these channels. Figures~\ref{fig_iso0} and \ref{fig_iso0_II} show the isospin 0 PB,VB amplitudes for $g_{KR} = 0, 1.5, 3$ and $6$ and Tables~\ref{iso0_1}-\ref{iso0_6} show the poles found by changing  $g_{KR}$ gradually from 0 to 6 in steps of 1. It is worth discussing these poles one by one.

\subsubsection{Two poles of the $\Lambda(1405)$ resonance:}
As shown in Tables~\ref{iso0_1} and \ref{iso0_2}, we find two poles for the $\Lambda(1405)$ resonance when $g_{KR} = 0$ at:
$1377 - i 63$ MeV and $1430 - i 15$ MeV. The former one couples strongly to $\pi \Sigma$ and the latter one to  $\bar{K} N$ 
in agreement with the findings of Refs.~\cite{jido1,jido2}. 
{\squeezetable
\begin{table}[htbp]
\caption[]{ $g^{i}$ couplings for  the lower mass pole of the  $\Lambda(1405)$ to PB and VB channels for different strengths of  the coupling between PB-VB systems. Note that  the couplings have been listed with
a precision to the first decimal place in this (and subsequent) table(s), which  leads to rounding off of a number smaller than 0.05 to 0.0 for $g_{KR} > 0$.}  \label{iso0_1}
\begin{ruledtabular}
\begin{tabular}{lrrrrrrr}
PB-VB &&&&&&&\\
coupling:  $g_{KR}$&\multicolumn{1}{c}{0}&\multicolumn{1}{c}{1}&\multicolumn{1}{c}{2}&\multicolumn{1}{c}{3}&\multicolumn{1}{c}{4}&\multicolumn{1}{c}{5}&\multicolumn{1}{c}{6}\\
\hline
$M_R - i\Gamma/2$ (MeV) $\longrightarrow$ &$1377 - i63$ &$1376 - i63$&$1374 - i62$&$1372 - i61$&$1368 - i59$&$1363 - i56$&$1357 - i53$\\
\hline
Channels $\downarrow$& \multicolumn{7}{c}{Couplings ($g^i$) of the poles to the different channels} \\
\hline
$\bar{K} N$             &$~1.4 - i 1.6$ &$~1.4 - i 1.6$&$~1.4 - i 1.6$&$~1.3 - i 1.6$&$~1.3 - i 1.5$&$~1.2 - i 1.5$&$~1.1 - i 1.4$\\
$\pi \Sigma$           &$-2.3 + i 1.4$ &$-2.3 + i 1.5$&$-2.2 + i 1.5$&$-2.3 + i 1.4$&$-2.3 + i 1.4$&$-2.2 + i 1.4$&$-2.2 + i 1.4$\\
$\eta \Lambda$      &$~0.2 - i 0.7$ &$~0.2 - i 0.6$&$~0.2 - i 0.6$&$~0.1 - i 0.6$&$~0.1 - i 0.6$&$~0.1 - i 0.6$&$~0.1 - i 0.6$\\
$K \Xi$                      &$-0.4 + i 0.4$ &$-0.4 + i 0.4$&$-0.4 + i 0.4$&$-0.5 + i 0.4$&$-0.5 + i 0.4$&$-0.5 + i 0.4$& $-0.6 + i 0.4$\\
$\bar{K^*}N$            &$~0.0 + i 0.0$ &$-0.4 + i 0.1$&$-0.7 + i 0.2$&$-1.1 + i 0.3$&$-1.4 + i 0.5$&$-1.6 + i 0.5$&$-1.7 + i 0.7$\\
$\omega \Lambda$&$~0.0 + i 0.0$ &$-0.2 - i 0.1$&$-0.3 - i 0.1$&$-0.4 - i 0.2$&$-0.6 - i 0.2$&$-0.7 - i 0.3$&$-0.7 - i 0.3$\\
$\rho \Sigma$          & $~ 0.0 + i 0.0$&$~0.1 + i 1.2$&$~0.3 + i 2.3$&$~0.4 + i 3.6$&$~0.7 + i 4.7$&$~0.9 + i 5.7$&$~1.3 + i 6.8$\\
$\phi \Lambda$       &$~0.0 + i 0.0$ &$~0.2 + i 0.1$&$~0.4 + i 0.2$&$~0.6 + i 0.3$&$~0.8 + i 0.3$&$~0.9 + i 0.4$&$~1.0 + i 0.5$\\
$K^*\Xi$                    & $~0.0 + i 0.0$&$~0.2 + i 1.0$&$~0.4 + i 2.0$&$~0.5 + i 3.0$&$~0.7 + i 3.9$&$~0.9 + i 4.8$&$~1.3 + i 5.7$\\
\end{tabular}
\end{ruledtabular}
\end{table}}

{\squeezetable
\begin{table}[htbp]
\caption[]{$g^{i}$ couplings for the higher mass pole of the $\Lambda(1405)$. 
} \label{iso0_2}
\begin{ruledtabular}
\begin{tabular}{lrrrrrrr}
PB-VB &&&&&&&\\
coupling:  $g_{KR}$&\multicolumn{1}{c}{0}&\multicolumn{1}{c}{1}&\multicolumn{1}{c}{2}&\multicolumn{1}{c}{3}&\multicolumn{1}{c}{4}&\multicolumn{1}{c}{5}&\multicolumn{1}{c}{6}\\
\hline
$M_R - i\Gamma/2$ (MeV) $\longrightarrow$ &$1430 - i15$ &$1430 - i15$&$1428 - i15$&$1426 - i14$&$1422 - i14$&$1418 - i12$&$1412 - i11$\\
\hline
Channels $\downarrow$& \multicolumn{7}{c}{Couplings ($g^i$) of the poles to the different channels} \\
\hline
$\bar{K} N$              &$~2.4 + i 1.1$ &$~2.4 + i 1.1$&$~2.4 + i 1.0$&$~2.5 + i 0.9$&$~2.6 + i 0.8$&$~2.7 + i 0.7$&$~2.8 + i 0.5$\\
$\pi \Sigma$             &$-0.2 - i 1.4$ &$-0.2 - i 1.3$&$-0.2 - i 1.3$&$-0.2 - i 1.3$&$-0.2 - i 1.2$&$-0.2 - i 1.2$ &$-0.2 - i 1.1$\\
$\eta \Lambda$       & $~1.3 + i 0.3$&$~1.4 + i 0.3$&$~1.4 + i 0.3$&$~1.4 + i 0.2$&$~1.4 + i 0.2$&$~1.5 + i 0.1$&$~1.5 + i 0.1$\\
$K \Xi$                      &$~0.0 - i 0.3$ &$~0.0 - i 0.3$&$~0.0 - i 0.3$&$~0.0 - i 0.3$&$~0.0 - i 0.3$&$~0.0 - i 0.3$&$~0.0 - i 0.3$ \\
$\bar{K^*}N$           &$~0.0 + i 0.0$ &$~0.1 - i 0.9$&$~0.1 - i 1.8$ &$~0.2 - i 2.7$&$~0.1 - i 3.6$&$~0.0 - i 4.5$&$-0.1 - i 5.3$\\
$\omega \Lambda$&$~0.0 + i 0.0$ &$~0.1 - i 0.3$&$~0.1 - i 0.6$&$~0.2 - i 0.9$&$~0.2 - i 1.2$&$~0.2 - i 1.5$&$~0.2 - i 1.8$\\
$\rho \Sigma$          &$~0.0 + i 0.0$ &$-0.5 - i 0.3$&$-0.9 - i 0.6$&$-1.3 - i 0.9$&$-1.7 - i 1.2$&$-2.1 - i 1.4$&$-2.4 - i 1.6$\\
$\phi \Lambda$       &$~0.0 + i 0.0$ &$-0.1 + i 0.4$&$-0.2 + i 0.9$&$-0.2 + i 1.3$&$-0.3 + i 1.7$&$-0.3 + i 2.2$&$-0.3 + i 2.6$\\
$K^*\Xi$                    &$~0.0 + i 0.0$ &$-0.4 - i 0.1$&$-0.8 - i 0.3$&$-1.1 - i 0.4$&$-1.5 - i 0.5$&$-1.7 - i 0.5$&$-2.0 - i 0.5$\\
\end{tabular}
\end{ruledtabular}
\end{table}}

By switching on the coupling between PB-VB channels, but keeping it small, i.e., $g_{KR}= 1$, we see that the couplings remains very similar for the PB channels while a small coupling for the VB channels develops. On increasing $g_{KR}$ merely to 2, we find that the coupling of the lower pole to the closed VB channels $\rho \Sigma$ and $K^* \Xi$ becomes similar to those of the $\bar{K} N$ and $\pi \Sigma$ channels  and the coupling of the higher pole ($1430 - i 15$ MeV) to $\bar{K}^* N$ becomes comparable to that of $\bar{K} N$. If $g_{KR} $ is fixed to half of its strength (which is 3), the couplings of the two $\Lambda(1405)$ poles  to the PB channels  remains almost unchanged but increase for the VB channels.

The full strength of $g_{KR} $ leads to a slight shift in the masses and widths of the two poles, we find them at  $1363 - i 56$ MeV and $1412 - i 11$ MeV which couple much stronger to some VB channels than to PB. The coupling of the pole at $1363 - i 56$ MeV to $\rho \Sigma$ and $K^* \Xi$ is twice 
the one to $\pi \Sigma$ while the latter coupling is very similar to the one obtained for $g_{KR} = 0$. The pole at $1412 - i 11$ MeV  turns out to couple strongly to the $\bar{K}^* N$ channel, almost two times more than to the $\bar{K} N$ channel.  Interestingly, since neither the pole positions nor the coupling of the $\Lambda(1405)$ poles to PB get much altered by the inclusion of VB as coupled channels, the amplitudes for the PB channels on the real energy axis continue looking very similar, except for a change in the strength, as shown in Fig.~\ref{fig_iso0} (left column). The amplitude for the  $\pi \Sigma$ channel, for which data is available, changes only slightly. Though it depicts one curious feature, that is a zero at the mass of the higher pole at $g_{KR} =$ 6, unlike the amplitude calculated with  $g_{KR} =$ 0.
Our results  show that although the PB channels can generate the $\Lambda(1405)$ and the available data  can be explained with this information, a better understanding of the structure of the $\Lambda(1405)$ requires the consideration of the VB channels. 
The (diagonal) amplitudes for the VB channels are also shown in  Fig.~\ref{fig_iso0} (right column), where the presence of the two poles of the  $\Lambda (1405)$ can be seen.

\subsubsection{$\Lambda(1670)$:}
A resonance around 1700 MeV, which was interpreted as a  $K \Xi$ bound state, was found in a study of PB systems \cite{bennhold}. The pole associated to this resonance was found to be very sensitive to the subtraction constant for the $K\Xi$  channel ($a_{K\Xi}$), it appeared at $1680 - i 20$ MeV with $a_{K\Xi} = -$ 2.67  and at $1708 - i21$ MeV with  $a_{K\Xi} = - $ 2.52.  In our case, using the same cut-off's throughout our study ( $\Lambda_{VB} =$ 545 MeV for the VB systems and  $\Lambda_{PB} =$ 750 MeV for PB channels) we find an isoscalar  pole at  $1767 - i25$ MeV when PB and VB are not coupled (keeping $g_{KR} =$  0). This pole  couples strongly to the $K\Xi$ channel,  as can be seen in Table~\ref{iso0_3}.
On allowing the PB-VB systems to couple, we find that the coupling of this pole to the PB channels remains almost unchanged (see Table~\ref{iso0_3}).  
{\squeezetable
\begin{table}[htbp]
\caption[]{$g^{i}$ couplings for the pole related to $\Lambda(1670)$. } \label{iso0_3}
\begin{ruledtabular}
\begin{tabular}{lrrrrrrrrrr}
PB-VB &&&&&&&\\
coupling:  $g_{KR}$&\multicolumn{1}{c}{0}&\multicolumn{1}{c}{1}&\multicolumn{1}{c}{2}&\multicolumn{1}{c}{3}&\multicolumn{1}{c}{4}&\multicolumn{1}{c}{5}&\multicolumn{1}{c}{6}\\
\hline
$M_R - i\Gamma/2$ (MeV) $\longrightarrow$ & $1767 - i25$&$1766 - i25$&$1763 - i25$&$1759 - i25$&$1754 - i26$&$1749 - i27$&$1744 - i28$\\
\hline
Channels $\downarrow$& \multicolumn{7}{c}{Couplings ($g^i$) of the poles to the different channels} \\
\hline
$\bar{K} N$             & $~0.2 - i 0.5$&$~0.2 - i 0.5$&$~0.2 - i 0.6$&$~0.2 - i 0.6$&$~0.2 - i 0.6$&$~0.2 - i 0.6$&$~0.3 - i 0.6$\\
$\pi \Sigma$           & $~0.1 + i 0.2$&$~0.1 + i 0.2$&$~0.1 + i 0.2$&$~0.1 + i 0.2$&$~0.1 + i 0.3$&$~0.1 + i 0.3$&$~0.1 + i 0.3$\\
$\eta \Lambda$      &$-1.0 + i 0.3$ &$-1.0 + i 0.3$&$-1.0 + i 0.3$&$-1.0 + i 0.3$&$-1.0 + i 0.3$&$-1.0 + i 0.3$&$-1.0 + i 0.3$\\
$K \Xi$                     &$~3.2 + i 0.3$ &$~3.2 + i 0.3$&$~3.2 + i 0.3$&$~3.2 + i 0.3$&$~3.3 + i 0.3$&$~3.3 + i 0.3$&$~3.4 + i 0.2$\\
$\bar{K^*}N$           &$~0.0 + i 0.0$ &$~0.0 + i 0.4$&$-0.1 + i 0.6$&$-0.2 + i 0.8$&$-0.2 + i 0.9$&$-0.3 + i 1.0$&$-0.3 + i 1.1$\\
$\omega \Lambda$&$~0.0 + i 0.0$ &$~0.0 + i 0.1$&$~0.1 + i 0.1$&$~0.1 + i 0.1$&$~0.1 + i 0.1$&$~0.1 + i 0.0$&$~0.1 - i 0.1$\\
$\rho \Sigma$         &$~0.0 + i 0.0$ &$~0.0 - i 0.8$&$~0.0 - i 1.6$&$~0.0 - i 2.2$&$~0.1 - i 2.8$&$~0.2 - i 3.2$&$~0.3 - i 3.5$\\
$\phi \Lambda$      &$~0.0 + i 0.0$ &$-0.1 - i 0.1$&$-0.1 - i 0.1$&$-0.1 - i 0.1$&$-0.1 - i 0.1$&$-0.1- i 0.0$&$-0.2 + i 0.1$\\
$K^*\Xi$                   &$~0.0 + i 0.0$ &$~0.0 - i 0.3$&$~0.1 - i 0.6$&$~0.2 - i 0.8$&$~0.3 - i 1.0$&$~0.4 - i 1.1$&$~0.5 - i 1.2$
\end{tabular}
\end{ruledtabular}
\end{table}}
However it develops a very strong coupling to the $\rho \Sigma$ channel.  The pole position is found to shift to lower energies while  $g_{KR}$ increases. It ends up at $1744 - i28$ MeV for $g_{KR} =$ 6. This value is still high as compared to the observed mass of the $\Lambda(1670)$ but we have shown  that the inclusion of VB coupled channels improves the agreement.  

The behavior of this pole can  be seen in the $K \Xi$  amplitudes shown in Fig.~\ref{fig_iso0}, which hardly change
with $g_{KR}$, and in the $\rho \Sigma$ amplitude in Fig.~\ref{fig_iso0_II} which shows a clear peak near 1740 MeV for $g_{KR}$ = 3, 6 (dash-dot and solid curves).

\subsubsection{States with higher masses (in the range of 1800-2100 MeV):}
The main objective of our present work  is to study the effect of the VB coupled channels on the low lying resonances. In order to 
draw any concrete conclusion about higher mass resonances, we should take into account a more complete VB $\rightarrow$ VB interactions as shown in Ref.~\cite{us}. However, within the present formalism, we can test  if the widths of the resonances in 1800-2100 MeV region increase a lot by coupling the lower mass (PB) open channels.

In this energy region, we find three (spin degenerate) poles in the isospin zero VB systems (with $g_{KR} =$ 0):  $1795 - i0 $ MeV, $1923 - i 4$ MeV and $2138 - i 21$ MeV,  quite in agreement with Ref.~\cite{eulogiopb}. The little differences in the pole positions found in our work and  Ref.~\cite{eulogiopb} arise due to the differences in the calculations of the loops. Firstly, we regularize the loops using a cut-off while the dimensional regularization scheme is used in Ref.~\cite{eulogiopb}.  We do not use the scheme of Ref.~\cite{eulogiopb} since, as discussed in Section~3.1, it gives rise to an inappropriate behavior of  the loops at energies much lower than the threshold. Secondly,  to consider the fact that the vector mesons can sometimes possess considerably large widths, as is the case of  $\rho$ and $K^*$, a convolution of the loops over the  $\rho$ and $K^*$ widths was made for the related channels in Ref.~\cite{eulogiopb} and, hence, the amplitudes and poles were obtained using the convoluted loops in the calculations. However, we restrict the calculation of the convoluted  loops to the real energy axis only.  This is so because the consideration of a varied  mass for the vector mesons implies the presence of a band of (meson-baryon) thresholds instead of a fixed value, which sometimes makes it difficult to look for poles moving in the complex plane.  This difficulty was also encountered by the authors of  Ref.~\cite{eulogiopb} although not in the case of systems with total isospin 0 and strangeness $-1$. 

We would also like to remind the reader that although the VB interaction obtained from the $t$-channel exchange gives rise to  spin independent amplitudes, the coupling of the PB and VB channels affects only the spin 1/2 amplitudes.  As a matter of course, in the present work we are only discussing the results obtained in the spin 1/2 case.  

To make the further discussion clearer, we will denote the (spin 1/2) poles found in the VB systems at   $1795 - i0 $ MeV, $1923 - i 4$ MeV and $2138 - i 21$ MeV as $P_{i}$, $P_{ii}$ and $P_{iii}$, respectively, since we cannot label them as $N^*$'s. This is so because a clear association of these poles to the known resonances cannot be made easily as the status of the known $N^*$ resonances in this energy region is very poor and a very little related information is available. Going back to the study of the poles, we find that the $P_{i}$, $P_{ii}$ poles  move to higher energies as the coupling between the PB and VB systems increases (see Tables~\ref{iso0_4} and  \ref{iso0_5}). 
{\squeezetable
\begin{table}[htbp]
\caption[]{$g^{i}$ couplings of the $P_{i}$ pole as a function of $g_{KR}$.} \label{iso0_4}
\begin{ruledtabular}
\begin{tabular}{lrrrrrcr}
PB-VB &&&&&&&\\
coupling:  $g_{KR}$&\multicolumn{1}{c}{0}&\multicolumn{1}{c}{1}&\multicolumn{1}{c}{2}&\multicolumn{1}{c}{3}&\multicolumn{1}{c}{4}&\multicolumn{1}{c}{~~~~5~~~~}&\multicolumn{1}{c}{6}\\
\hline
$M_R - i\Gamma/2$ (MeV) $\longrightarrow$ & $1795 - i 0$&$1797 - i 0.5$&$1802 - i 2$&$1812 - i 4$&$1822 - i 6.5$&-&$1844 - i94$\\
\hline
Channels $\downarrow$& \multicolumn{7}{c}{Couplings ($g^i$) of the poles to the different channels} \\
\hline
$\bar{K} N$              &$~0.0 + i 0.0$&$-0.1 + i 0.0$&  $-0.2 + i 0.1$&$-0.3 + i 0.1$&$-0.4 + i 0.0$&-&$-1.2 - i 0.5$\\
$\pi \Sigma$            &$~0.0 + i 0.0$&$-0.1 + i 0.0$&$-0.1 + i 0.1$&$-0.1 + i 0.1$&$-0.1 + i 0.0$&-&$-0.4 - i 0.3$\\
$\eta \Lambda$       &$~0.0 + i 0.0$&$-0.1 - i 0.0$&$-0.2 - i 0.1$&$-0.2 - i 0.1$&$-0.3 - i 0.1$&-&$-0.6 - i 0.2$\\
$K \Xi$                      &$~0.0 + i 0.0$&$~0.1 + i 0.2$&$~0.0 + i 0.3$&$~0.0 + i 0.4$&$~0.1 + i 0.3$&-&$~0.4 + i 0.3$ \\
$\bar{K^*}N$           &$~3.8 - i 0.0$&$~3.7 + i 0.0$& $~3.4 + i 0.1$&$~2.9 + i 0.3$&$~2.3 + i 0.5$&-&$~1.6 - i 0.8$\\
$\omega \Lambda$&$~1.2 - i 0.0$&$~1.2 + i 0.0$& $~1.1 + i 0.0$ &$~0.9 + i 0.1$&$~0.7 + i 0.1$&-&$~0.7 - i 0.1$\\
$\rho \Sigma$          &$-1.9 - i 0.0$&$-1.9 + i 0.0$&$-1.8 + i 0.1$&$-1.9 + i 0.2$&$-2.0 + i 0.1$&-&$-5.4 + i 0.0$\\
$\phi \Lambda$       &$-1.8 - i 0.0$&$-1.8 - i 0.0$&$-1.6 - i 0.1$&$-1.4 - i 0.1$ &$-1.1 - i 0.2$&-&$-1.2 + i 0.1$\\
$K^*\Xi$                    &$-0.6 - i 0.0$&$-0.5 + i 0.0$&$-0.5 + i 0.0$&$-0.5 + i 0.1$&$-0.5 + i 0.0$&-&$-1.2 - i 0.2$\\
\end{tabular}
\end{ruledtabular}
\end{table}
}
{\squeezetable
\begin{table}[htbp]
\caption[]{$g^{i}$ couplings for the $P_{ii}$ pole as a function of $g_{KR}$.  The sign $^{**}$ in this table indicates the unphysical nature of a pole.} \label{iso0_5}
\begin{ruledtabular}
\begin{tabular}{lrrrrccc}
PB-VB &&&&&&&\\
coupling:  $g_{KR}$&\multicolumn{1}{c}{0}&\multicolumn{1}{c}{1}&\multicolumn{1}{c}{2}&\multicolumn{1}{c}{3}&\multicolumn{1}{c}{4}&\multicolumn{1}{c}{5}&\multicolumn{1}{c}{6}\\
\hline
$M_R - i\Gamma/2$ (MeV) $\longrightarrow$ & $1923 - i 4$&$1926 - i 6$&$1934 - i10$&$1948 - i16$&$1969 - i11^{**}$&$2012 - i 6^{**}$&$2090 - i14^{**}$\\
\hline
Channels $\downarrow$& \multicolumn{7}{c}{Couplings ($g^i$) of the poles to the different channels} \\
\hline
$\bar{K} N$             &$~0.0 + i 0.0$&$-0.1 + i 0.1$&$-0.2 + i 0.1$ &$-0.3 + i 0.2$&- &-&-\\
$\pi \Sigma$            &$~0.0 + i 0.0$&$-0.1 + i 0.0$&$-0.2 + i 0.1$ &$-0.3 + i 0.1$&- &-&-\\
$\eta \Lambda$       &$~0.0 + i 0.0$&$~0.0 + i 0.0$&$~0.0 + i 0.0$&$~0.0 + i 0.1$&- &-&-\\
$K \Xi$                      &$~0.0 + i 0.0$&$-0.1 - i 0.0$&$-0.2 - i 0.1$&$-0.3 - i 0.3$&- &-&-\\
$\bar{K^*}N$            &$~0.1 - i 0.5$&$~0.0 - i 0.5$&$~0.0 - i 0.6$&$-0.1 - i 0.7$&- &-&-\\
$\omega \Lambda$&$~0.3 - i 0.2$&$~0.3 - i 0.2$&$~0.3 - i 0.2$&$~0.4 - i 0.3$&- &-&-\\
$\rho \Sigma$          &$~3.7 + i 0.2$&$~3.6 + i 0.2$&$~3.2 + i 0.4$&$~2.6 + i 0.9$&- &-&-\\
$\phi \Lambda$       &$-0.5 + i 0.3$&$-0.5 + i 0.3$&$-0.5 + i 0.3$&$-0.5 + i 0.4$&- &-&-\\
$K^*\Xi$                    &$~1.0 + i 0.0$&$~0.9 + i 0.0$&$~0.7 + i 0.1$&$~0.3 + i 0.3$&- &-&-\\
\end{tabular}
\end{ruledtabular}
\end{table}}
For small values of $g_{KR}$ (0-3), these poles are found to couple strongly to $\bar{K}^* N$ and $\rho \Sigma$, respectively. For  $g_{KR} =$ 4, we find the coupling of the $P_i$ pole to the $\rho \Sigma$ and $\bar{K}^* N$ channel becomes comparable  and the $P_{ii}$ pole becomes unphysical ($\rho \Sigma$ virtual state). For $g_{KR} =$ 5, we do not find any physical pole in 1800-2000 MeV but for $g_{KR} =$ 6 we find one physical pole at  $1844 - i94$ MeV which is listed in Table~ \ref{iso0_4}, although it cannot be connected to any of the two poles we started with. The couplings given in Table~\ref{iso0_4} show that the pole found at  $1844 - i94$ MeV can be interpreted as a $\rho \Sigma$ bound state.

The behavior of the poles $P_{i}$ and  $P_{ii}$ can also be seen in the amplitudes shown in Fig.~\ref{fig_iso0_II}, most clearly in  the $\rho \Sigma$ channel. The result corresponding to  $g_{KR} =$ 3 (dash dot line) shows three peaks; one for the $\Lambda (1670) $ and other  two related to the $P_{i}$ and  $P_{ii}$ poles. Further, the solid line, which corresponds to the full PB-VB coupling, shows a peak near $1844$ MeV.

The third pole found in the VB systems, in the 1800-2100 MeV region, denoted as $P_{iii}$, is found to couple strongly to $K^* \Xi$ (see Table~\ref{iso0_6}). 
\begingroup
\squeezetable
\begin{table}[htbp]
\caption[]{$g^{i}$ couplings for the $P_{iii}$ pole as a function of $g_{KR}$. } \label{iso0_6}
\begin{ruledtabular}
\begin{tabular}{lrrrrrrrrrrrr}
PB-VB&&&&&&\multicolumn{1}{c}{}&\multicolumn{2}{c}{}&$\quad$\\
coupling:  $g_{KR}$&\multicolumn{1}{c}{0}&\multicolumn{1}{c}{1}&\multicolumn{1}{c}{2}&\multicolumn{1}{c}{3}&\multicolumn{1}{c}{4}&\multicolumn{1}{c}{5}&\multicolumn{2}{c}{6}&$\quad\quad$\\
&&&&&&\multicolumn{1}{c}{}&\multicolumn{2}{c}{$\overbrace{\quad\quad\quad\quad\quad\quad\quad\quad}$}&$\quad\quad$\\
\cline{1-9}
$M_R - i\Gamma/2$ (MeV) $\longrightarrow$ & $2138 - i21$&$2138 - i21$&$ 2140 - i22$&$2143 - i23$&$2149 - i25$& $2159 - i33$ & $2151 - i119$ & $2160 - i73$&\\
\cline{1-9}
Channels $\downarrow$& \multicolumn{8}{c}{Couplings ($g^i$) of the poles to the different channels} &\\
\cline{1-9}
$\bar{K} N$             &$ 0.0 + i 0.0$ &$ 0.0 + i 0.0$&$ 0.0 + i 0.0$&$-0.1 + i 0.0$&$-0.1 + i 0.0$& $-0.1 + i 0.2$&$ 1.3 - i 0.3$& $ 0.2 + i 0.9$&\\
$\pi \Sigma$            & $ 0.0 + i 0.0$&$ 0.0 + i 0.0$&$ -0.1 + i 0.0$&$-0.1 + i 0.1$&$-0.2 + i 0.1$&$-0.3 + i 0.2$ &$ 1.0 + i 0.1$& $-0.2 + i 0.8$&\\
$\eta \Lambda$       &$ 0.0 + i 0.0$ &$ 0.0 + i 0.0$&$ 0.0 + i 0.0$&$ 0.0 + i 0.1$&$ 0.0 + i 0.1$& $ 0.0 + i 0.2$&$ 0.7 + i 0.0$& $ 0.0 + i 0.6$&\\
$K \Xi$                      &$ 0.0 + i 0.0$ &$ 0.0 + i 0.0$&$ 0.0 + i 0.0$&$ 0.0 + i 0.0$&$ -0.1 + i 0.1$& $-0.1 + i 0.1$&$ 0.9 + i 0.3$&  $-0.3 + i 0.7$&\\
$\bar{K^*}N$            &$ 0.0 - i 0.4$ &$ 0.0 - i 0.5$&$ -0.1 - i 0.5$&$-0.1 - i 0.5$&$-0.1 - i 0.6$& $-0.2 - i 0.6$&$-0.4 + i 1.3$& $-1.0 - i 0.6$&\\
$\omega \Lambda$&$-0.5 + i 0.3$ &$-0.5 + i 0.3$&$-0.6 + i 0.3$&$-0.6 + i 0.3$&$-0.6 + i 0.2$& $-0.6 + i 0.1$&$-0.2 + i 1.1$& $-1.2 - i 0.0$&\\
$\rho \Sigma$          &$ 0.1 + i 0.1$ &$ 0.1 + i 0.1$&$ 0.1 + i 0.1$&$ 0.1 + i 0.0$&$ 0.0 - i 0.1$& $-0.2 - i 0.2$&$-0.5 + i 1.7$& $-1.3 - i 0.4$&\\
$\phi \Lambda$       &$ 0.7 - i 0.4$ &$ 0.8 - i 0.4$&$ 0.9 - i 0.4$&$ 0.9 - i 0.4$&$ 0.9 - i 0.4$& $ 0.9 - i 0.2$&$ 0.5 - i 1.6$& $ 1.8 + i 0.1$&\\
$K^*\Xi$                   &$ 4.2 + i 0.2$ &$ 4.3 + i 0.2$&$ 4.9 + i 0.3$&$ 4.7 + i 0.3$&$ 4.5 + i 0.5$& $ 4.3 + i 1.1$&$ 5.4 - i 4.9$& $ 6.6 + i 4.1$&\\
\end{tabular}
\end{ruledtabular}
\end{table}
\endgroup
This pole is found to shift to higher energies as the coupling between PB and VB channels increases, until $g_{KR} =$ 6 when we find a double pole structure with the pole positions being  $2151 - i 119$ MeV and $2160 - i 73$ MeV.  Both poles are found to couple strongly to the $K^* \Xi$ channel but the former one appears to couple to PB channels slightly more than the latter one.

To summarize, we started with  three poles in the 1800-2100 MeV energy region in the VB systems uncoupled to PB: $1795 - i0 $ MeV, $1923 - i 4$ MeV and $2138 - i 21$ MeV and end up also with three poles in the PB-VB coupled systems:  $1844 - i 94$, $2151 - i 119$ MeV and $2160 - i 73$ MeV but the nature of the latter set of poles is different as compared to the former ones.

An interesting finding of our work is that the width of the poles does not change a lot when they are allowed to couple to more open channels. Although such a notion exists that a pole would get moderately modified by taking into account  those coupled channels  which consist of hadrons with the total mass much smaller than the mass of  the resonance, it has not been explicitly verified earlier. Our results show that the general notion may not be very far from the reality.

\subsection{Isospin =1}
\begin{figure}[ht!]
\begin{tabular}{cc}
\includegraphics[width=0.35\linewidth,height=3.8cm]{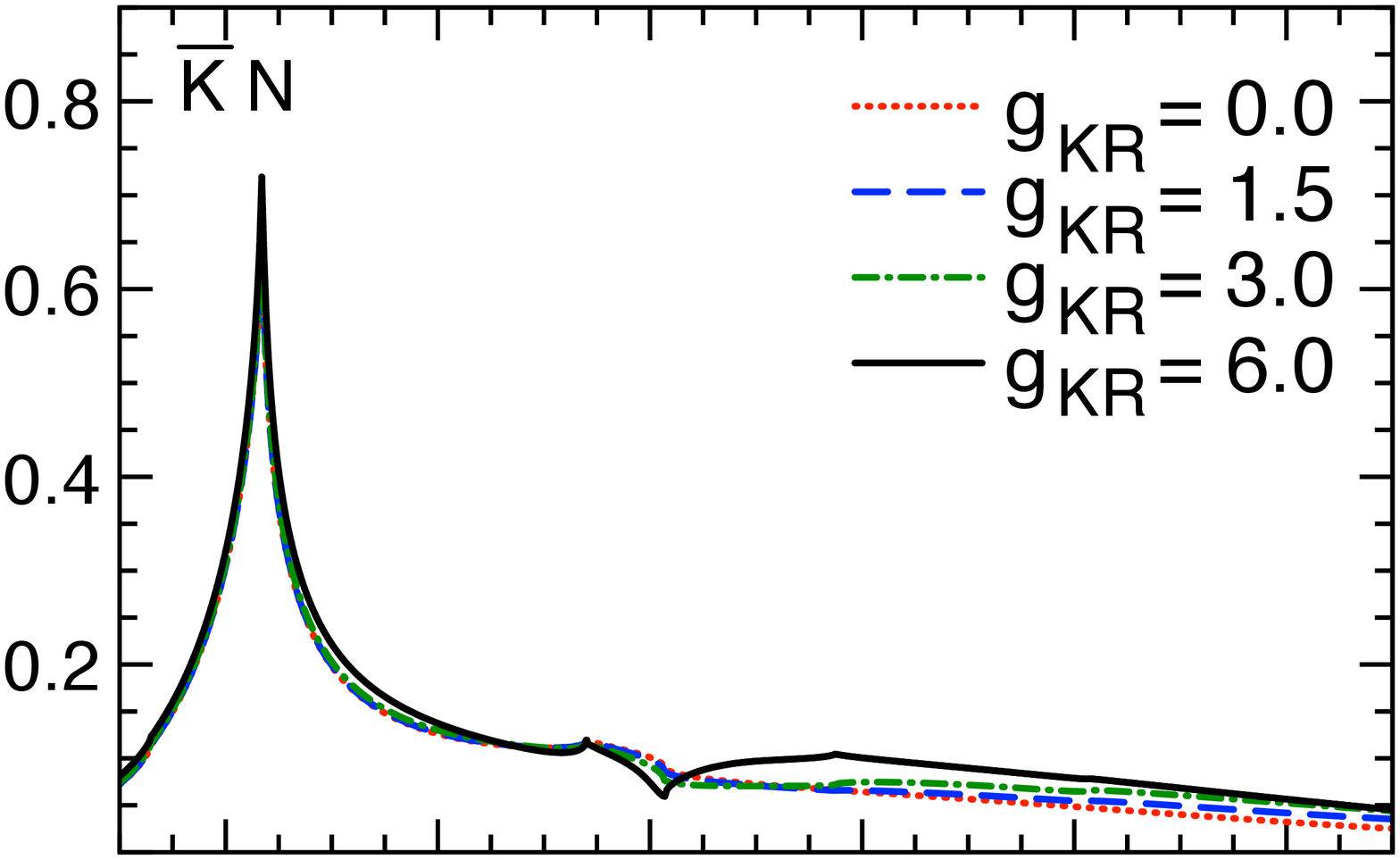}&\includegraphics[width=0.35\linewidth,height=3.8cm]{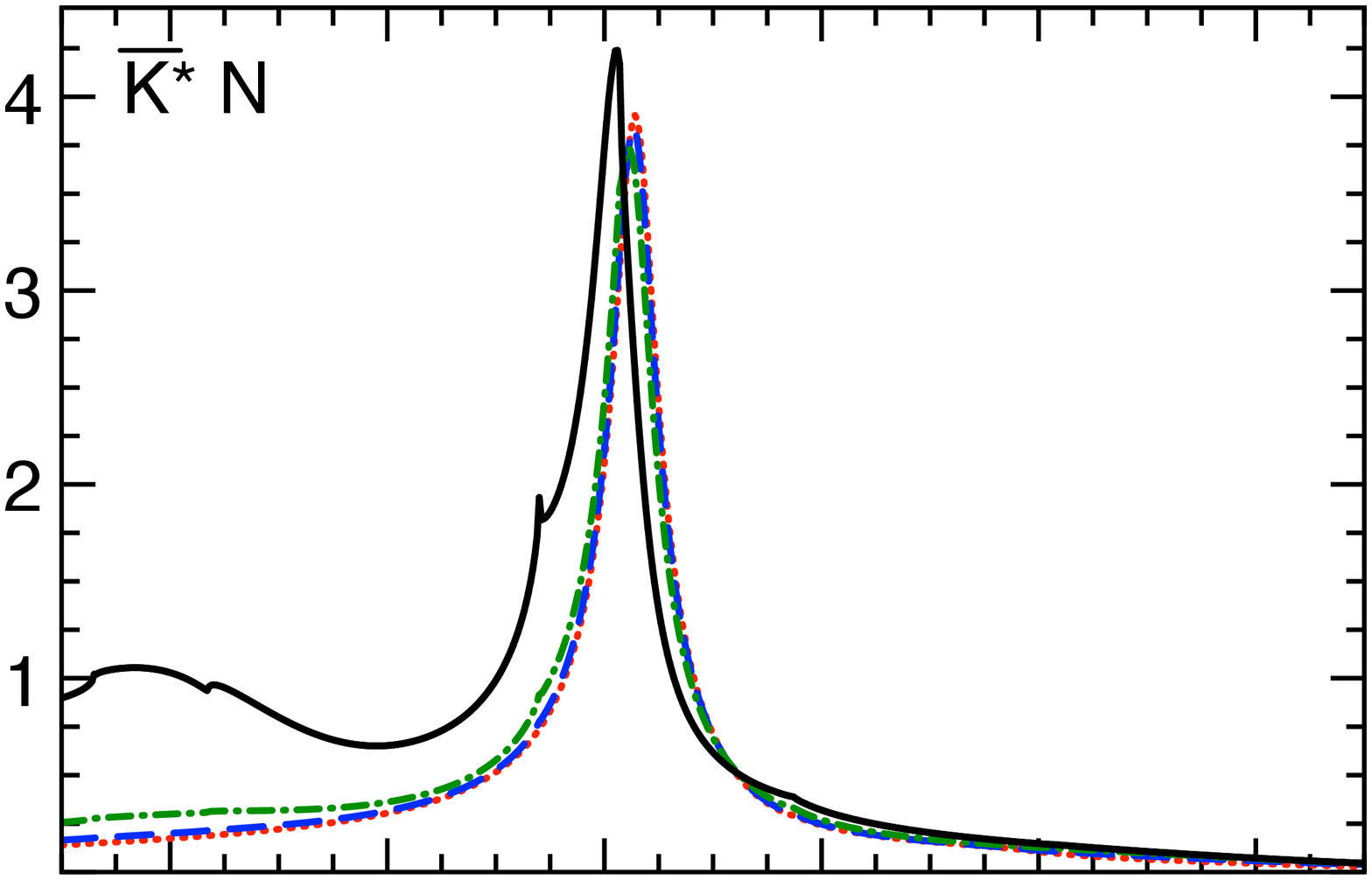}\\[-0.75cm]
\includegraphics[width=0.35\linewidth,height=3.8cm]{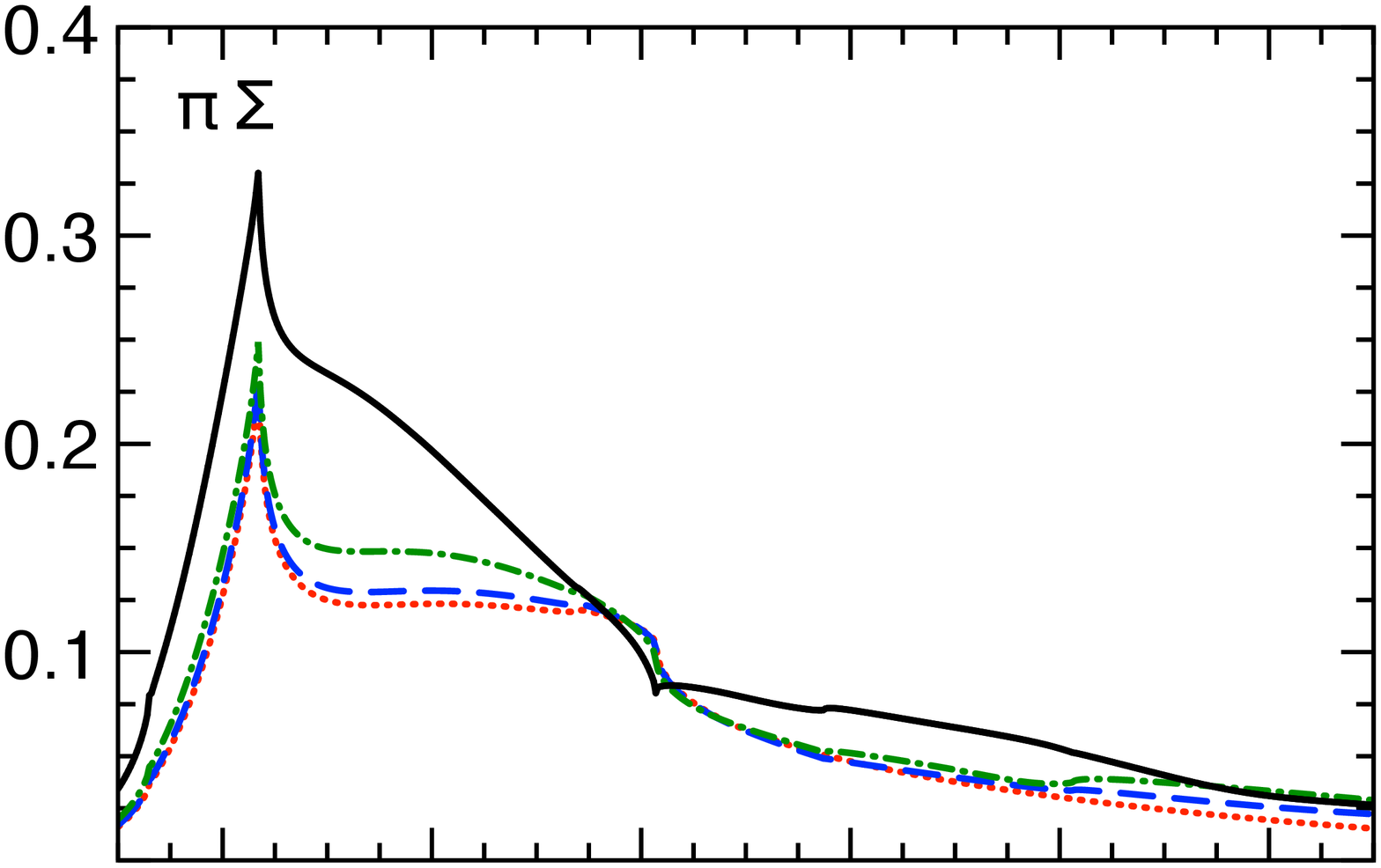}&\includegraphics[width=0.35\linewidth,height=3.8cm]{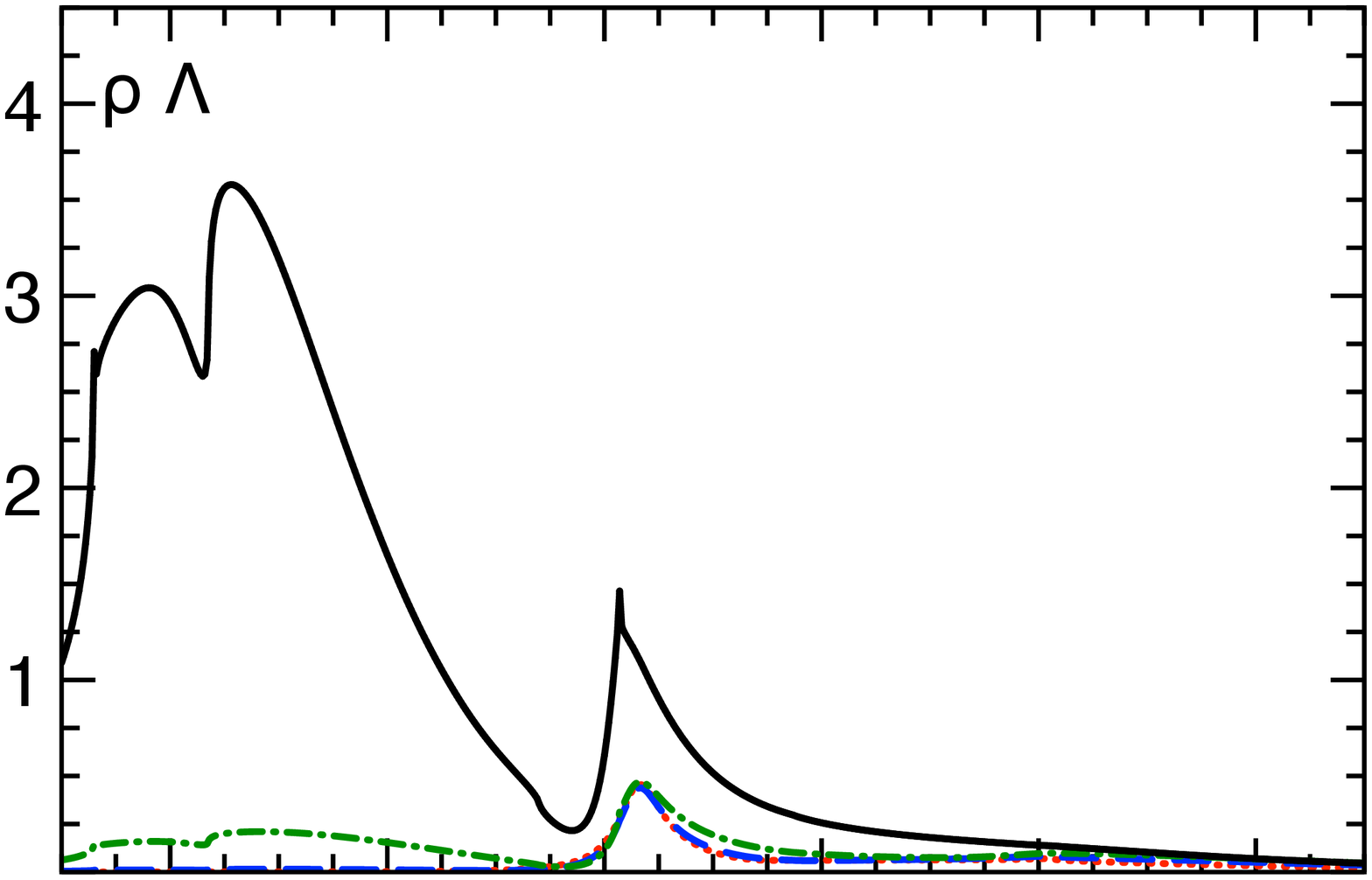}\\[-0.75cm]
\hspace{-0.89cm}\includegraphics[width=0.402\linewidth,height=3.8cm]{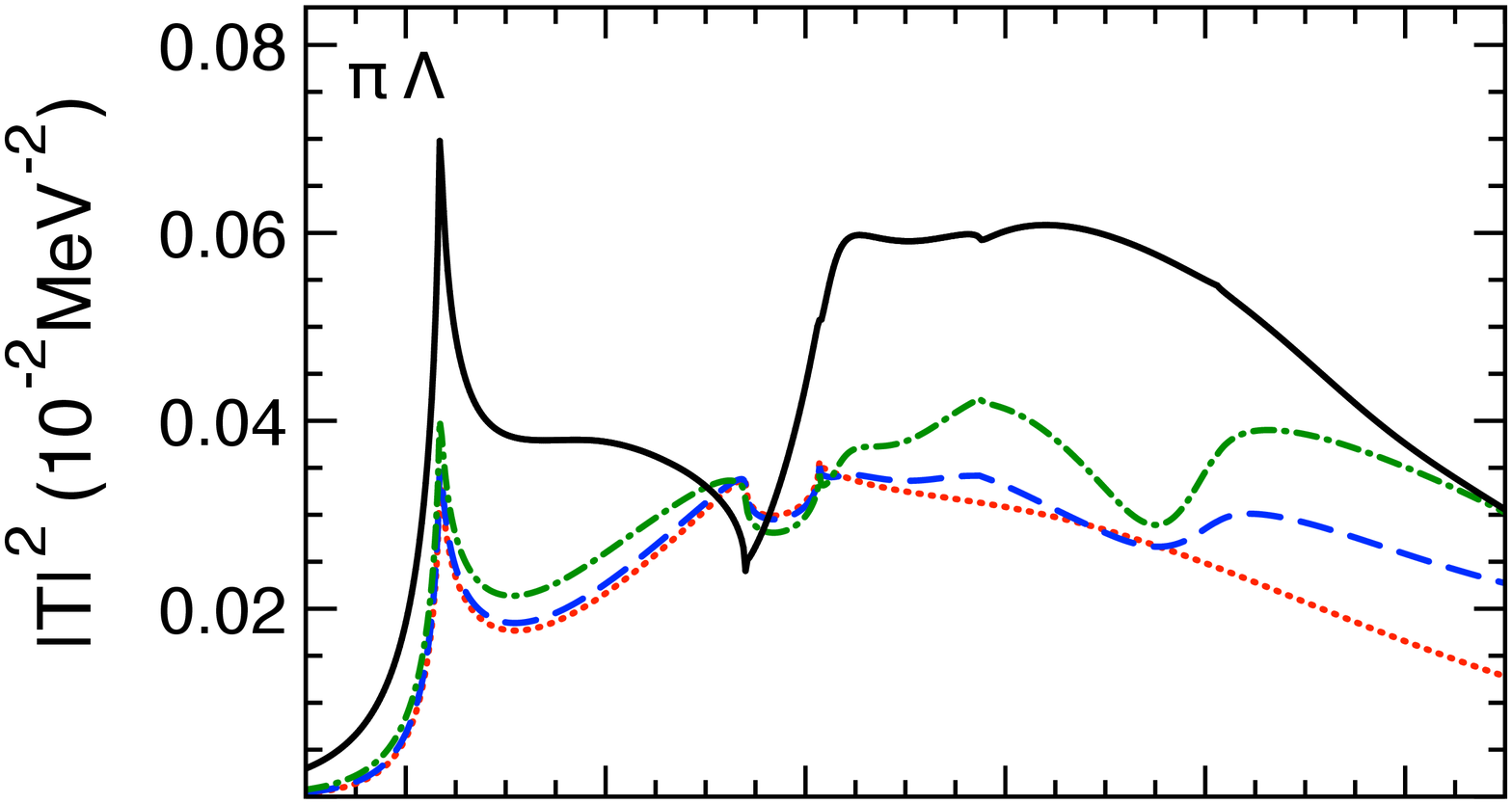}&\includegraphics[width=0.35\linewidth,height=3.8cm]{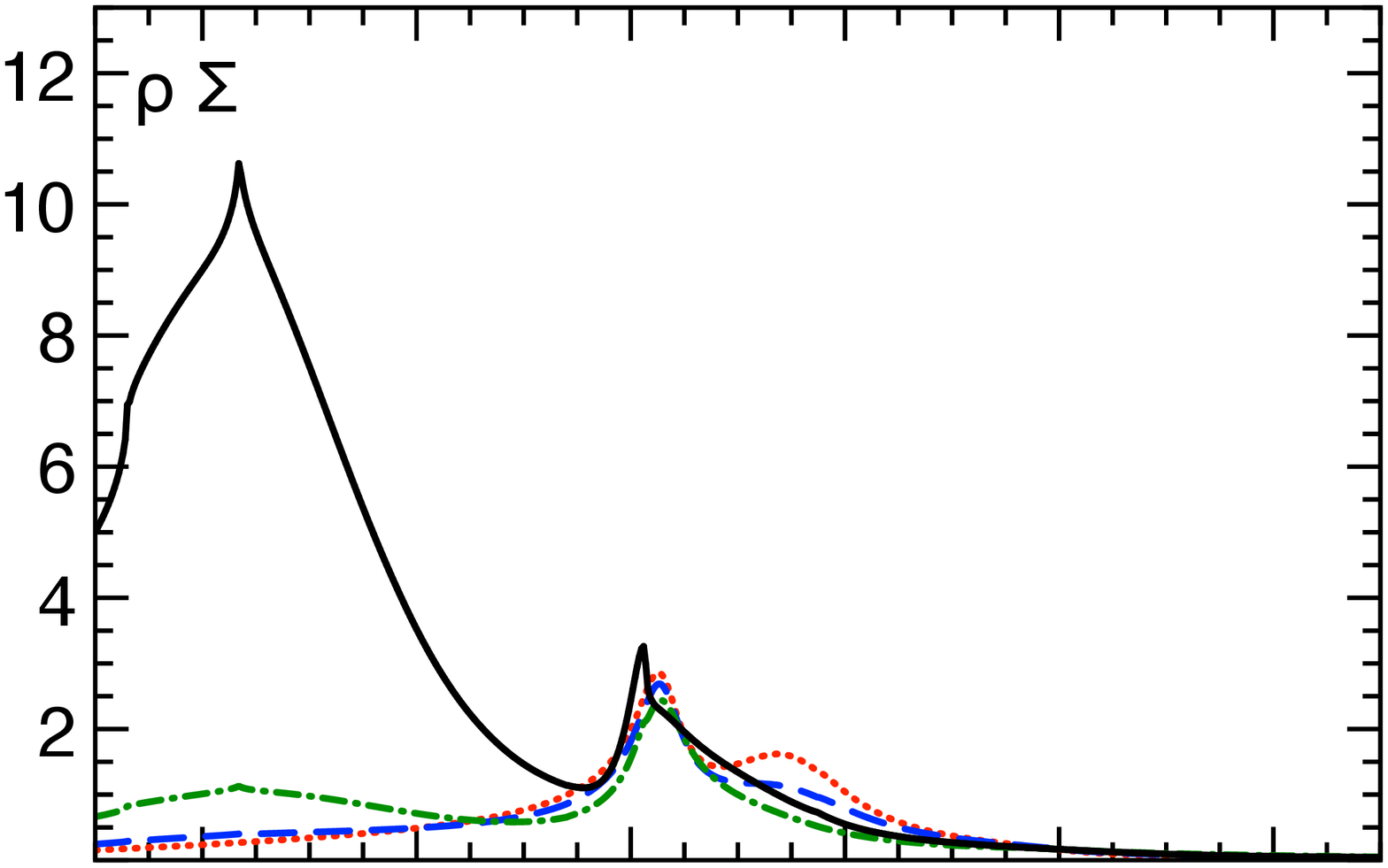}\\[-0.75cm]
\includegraphics[width=0.35\linewidth,height=3.8cm]{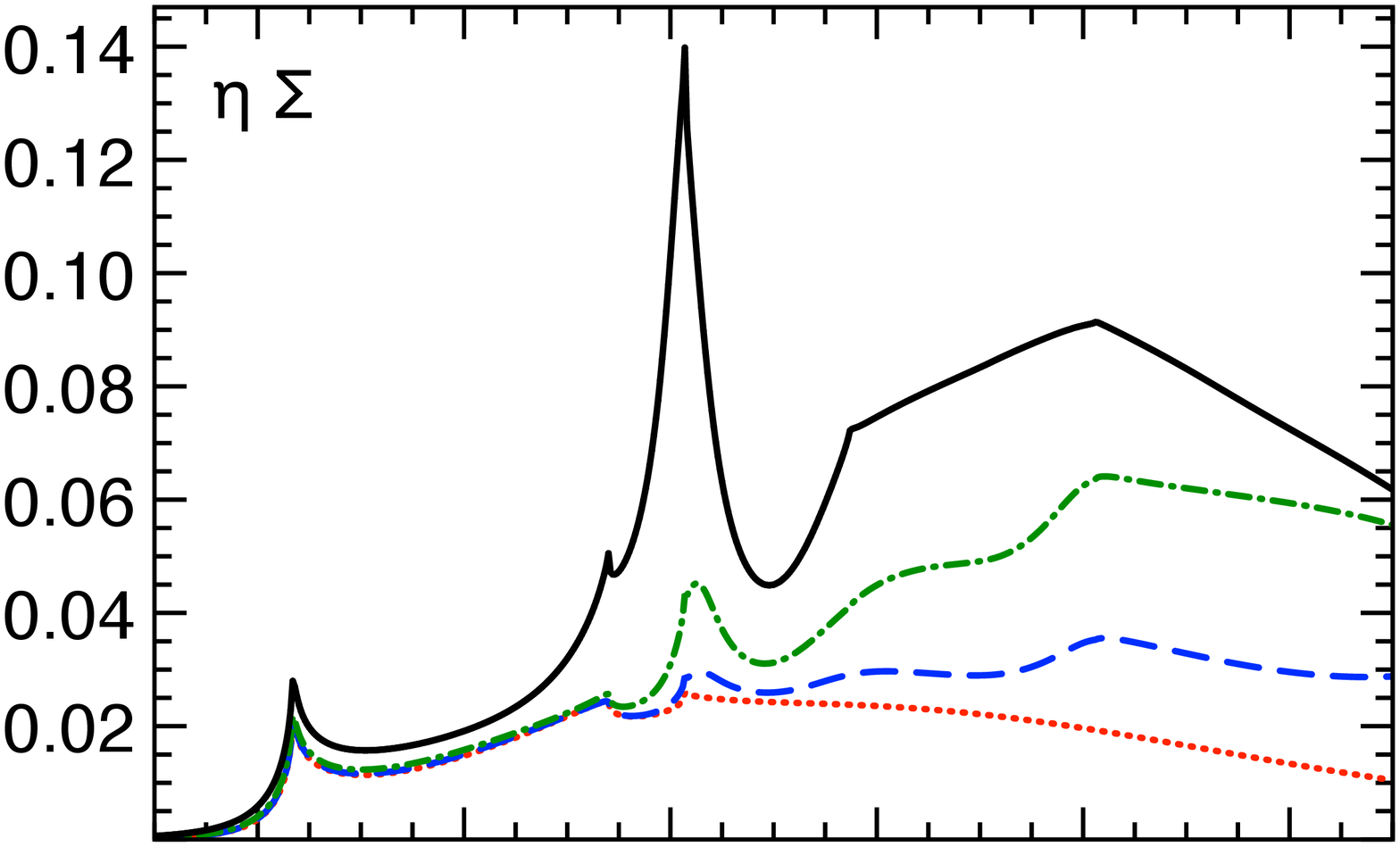}&\includegraphics[width=0.35\linewidth,height=3.8cm]{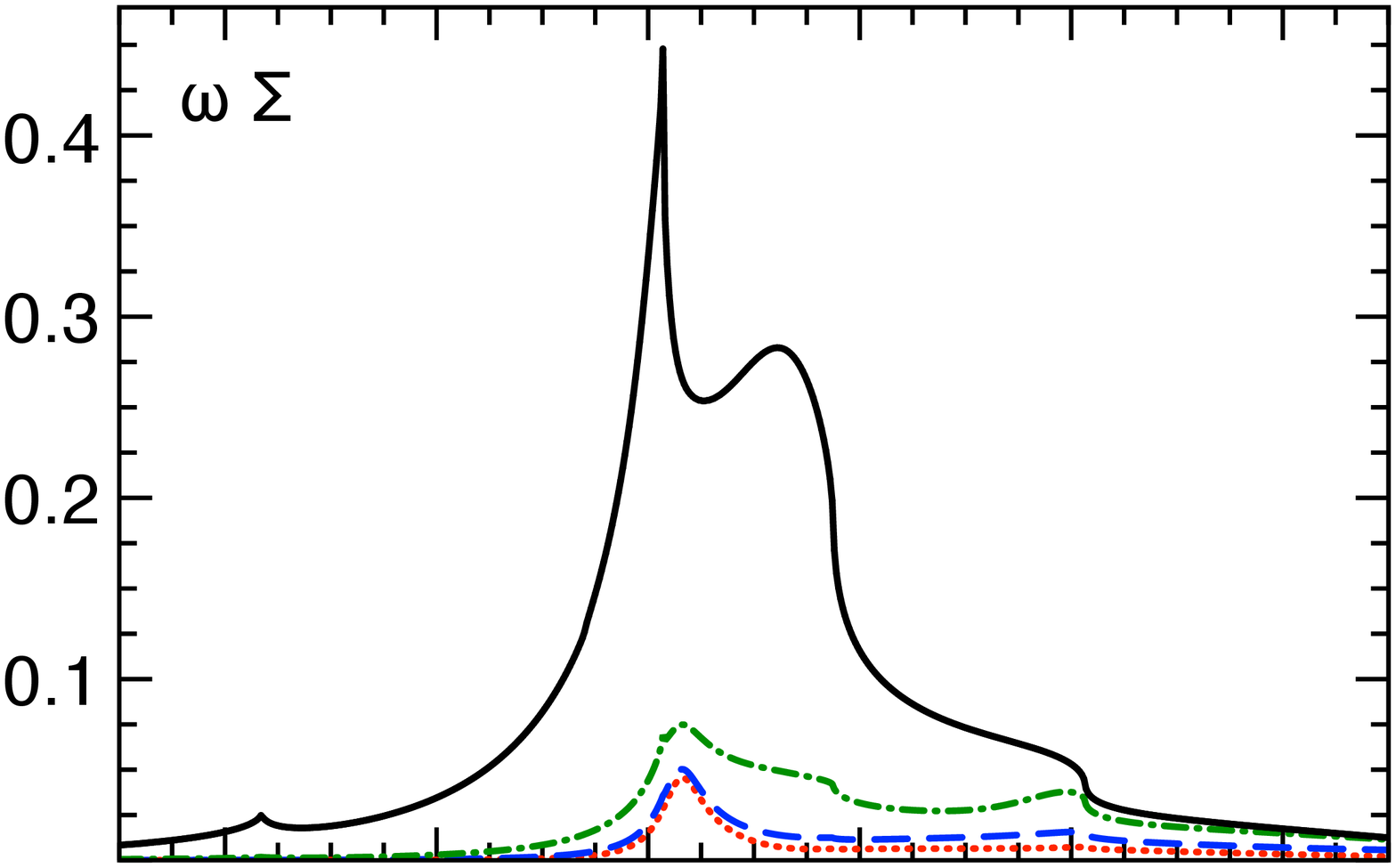}\\[-0.75cm]
\includegraphics[width=0.35\linewidth,height=4.6cm]{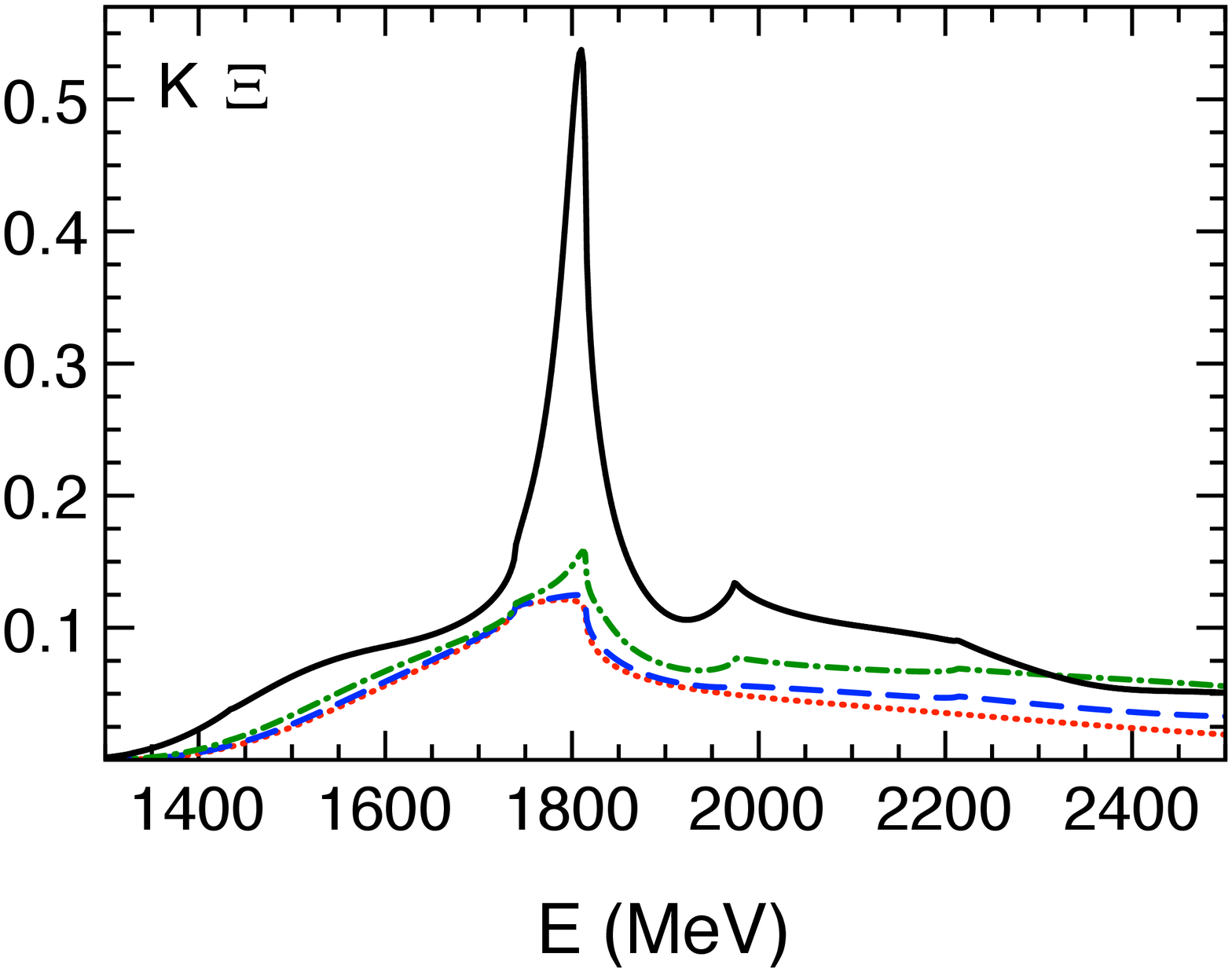}&\includegraphics[width=0.35\linewidth,height=4.6cm]{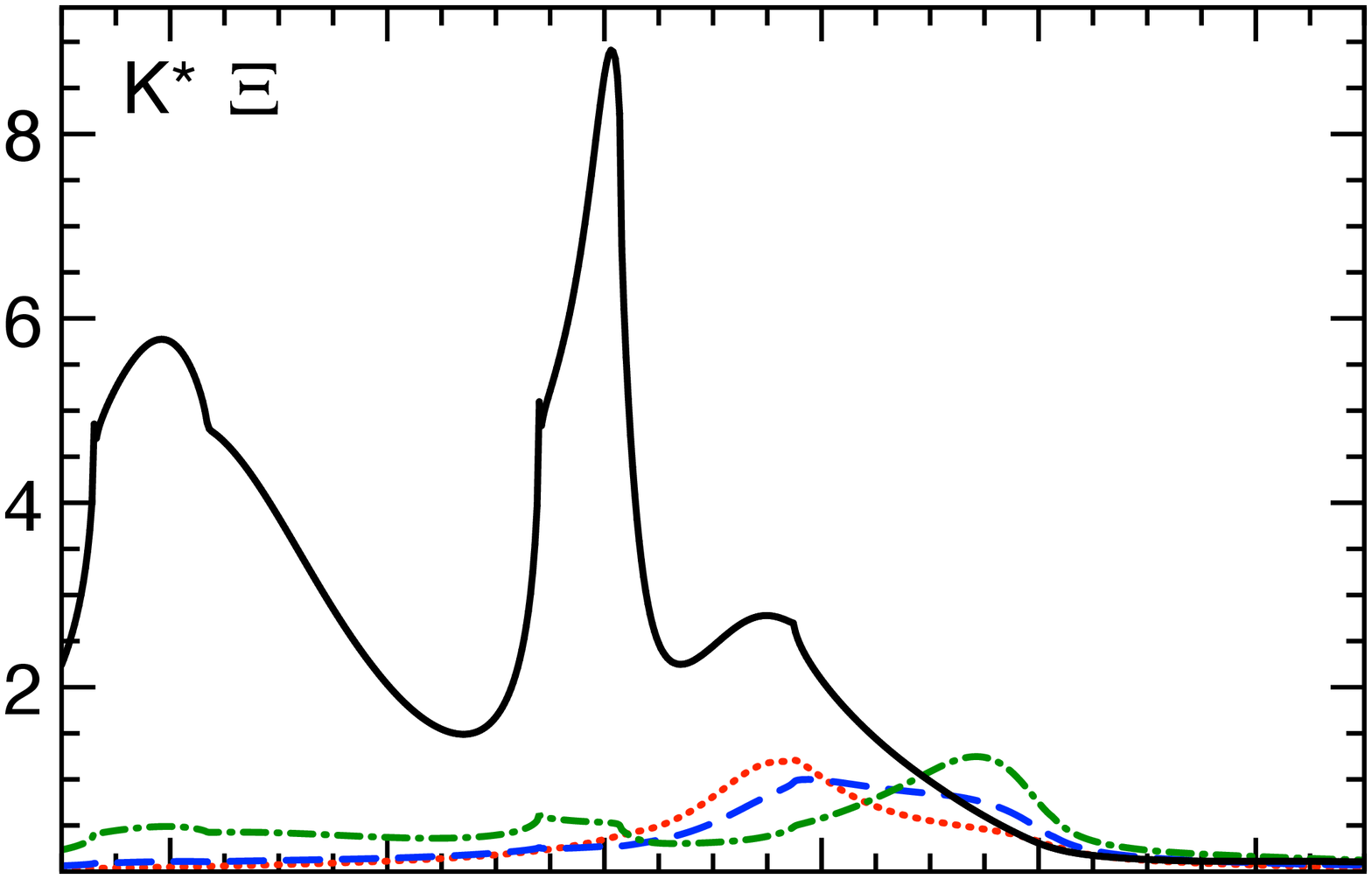}\\[-1.5cm]
&\includegraphics[width=0.35\linewidth,height=4.6cm]{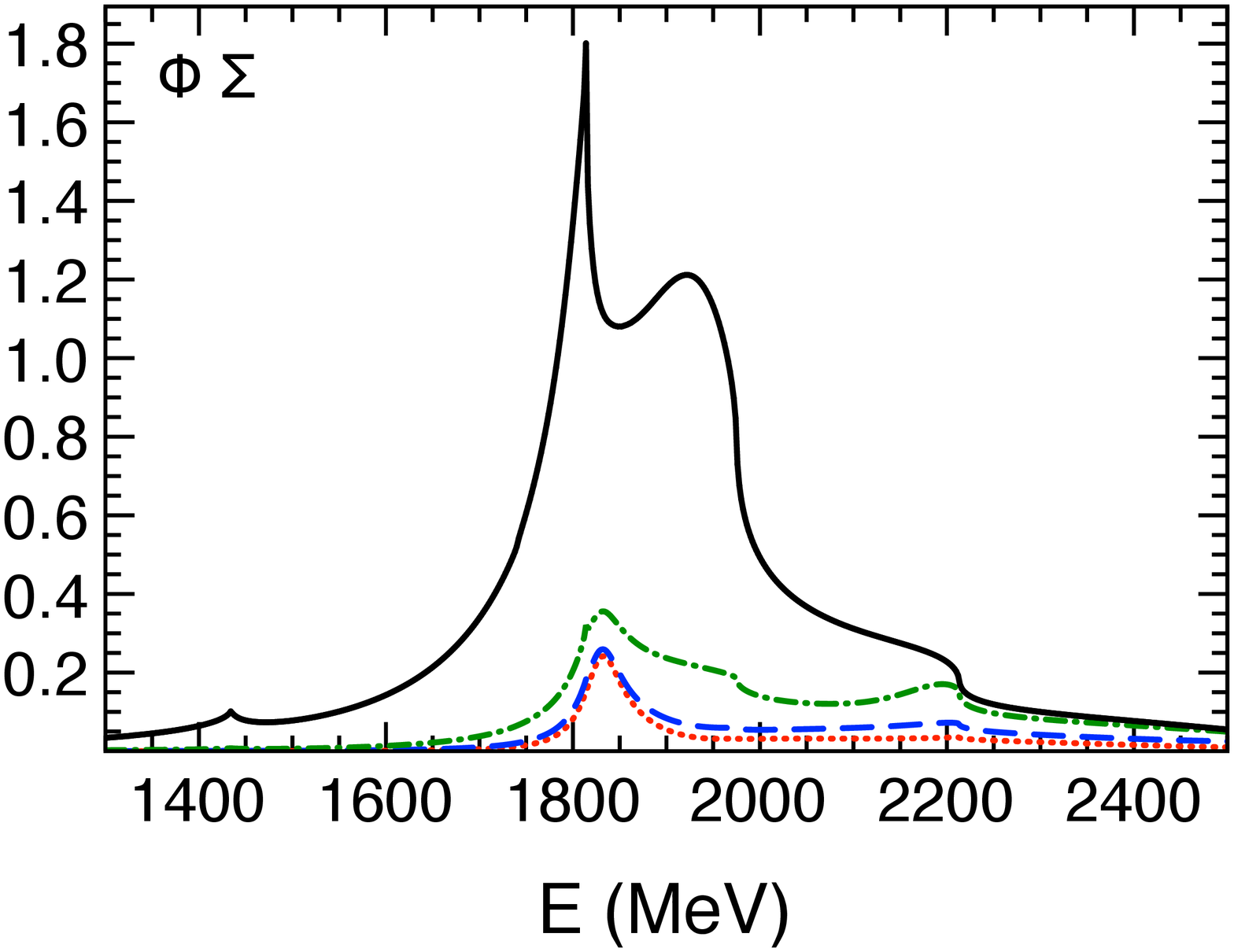}
\end{tabular}
\caption{Isospin one amplitudes of the PB and VB systems.}\label{fig_iso1}
\end{figure}

In the isospin 1 case, only one pole was found in the PB study of Ref.~\cite{bennhold} at $1579- i 264$ MeV which was related to the $1/2^-$  $\Sigma(1620)$
resonance, although  the width of the $\Sigma(1620)$ is $\leq$ 100 MeV.  Besides, this pole was found to be very sensitive to the subtraction constant parameters. Also, in the vector meson-baryon study of Ref.~\cite{eulogiovb}, two peaks were found in the amplitudes but no corresponding poles were found in the complex plane. It was explained in  Ref.~\cite{eulogiovb}  that finding poles was sometimes difficult due to the consideration of the widths of the vector mesons.  We find that even without this consideration, i.e., without convoluting the loops, only one physical pole appears in isospin 1 VB systems, with its position being very close to the $\bar{K}^* N$ threshold.  In view of the uncertain situation in the isospin 1  case, we do not try to adjust the cut-offs in our calculations to reproduce these poles. We, thus,  keep the same cut-offs  which we have used in the isospin 0 case, which reproduce the results of the previous PB studies. We show the amplitudes obtained in this way for the uncoupled PB and VB systems in the isospin 1 configuration as  dotted lines in Fig.~\ref{fig_iso1}. We find that these amplitudes are very similar to the ones obtained by calculating the loops with the dimensional regularization method as done in Refs.~\cite{eulogiopb,eulogiovb}. 

Corresponding to the amplitudes obtained with $g_{KR} =$ 0, we find a pole at $1479 -  i 285$ MeV in  the PB systems and another at $1831 -i 0$ in the VB systems. The pole obtained in the PB channels is found to couple  mostly to $K \Xi$ and the one in the VB system couples mostly to $\bar{K}^* N$. We show the couplings of these poles to the related channels in Table~\ref{iso1_1}.

\begin{table}[htbp]
\caption[]{$g^{i}$ couplings for the  poles found in the uncoupled PB and VB systems in the isospin 1 configuration.} \label{iso1_1}
\begin{center}
\begin{ruledtabular}
\begin{tabular}{lrrr}
$M_R - i\Gamma/2$ (MeV) $\longrightarrow$ & $1479 -  i 285 $ &$1831 - i0$ \\
\hline
Channels $\downarrow$& \multicolumn{2}{l}{Couplings ($g^i$) of the poles to the different channels} \\
\hline
$\bar{K} N$               &$ 0.4 - i 1.1$ & $ 0.0 + i 0.0$ \\
$\pi \Sigma$         &$ 1.4 - i 1.6$ &$ 0.0 + i 0.0$\\
$\pi \Lambda$          &$-0.6 + i 1.3$ &$ 0.0 + i 0.0$\\
$\eta \Sigma$        &$-0.5 + i 1.2$ &$ 0.0 + i 0.0$\\
$K \Xi$                      &$ 1.2 - i 2.4$ &$ 0.0 + i 0.0$\\
$\bar{K^*}N$           &$ 0.0 + i 0.0$&$ 1.0 - i 0.6$\\
$\rho  \Lambda$      &$ 0.0 + i 0.0$&$-0.7 + i 0.4$\\
$\rho  \Sigma$          &$ 0.0 + i 0.0$&$-0.7 + i 0.4$\\
$\omega  \Sigma$    &$ 0.0 + i 0.0$&$-0.4 + i 0.2$\\
$K^* \Xi$                    &$ 0.0 + i 0.0$&$ 0.1 - i 0.1$\\
$\phi  \Sigma$         &$ 0.0 + i 0.0$&$ 0.6 - i 0.3$\\
\end{tabular}
\end{ruledtabular}
\end{center}
\end{table}

We now discuss the  $T$-matrices calculated by coupling the PB and VB system with $g_{KR} =  6$. The amplitudes found with $g_{KR} =  6$ are shown
by  solid lines in  Fig.~\ref{fig_iso1}.  The calculation in the complex plane results in the finding of three poles at :  $1426 -  i143$ MeV,   $1439 -  i194$ MeV and  $2372 - i162$ MeV.  These poles and their couplings to different channels are listed in Table~\ref{iso1_2}. 
\begin{table}[h]
\caption[]{$g^{i}$ couplings  for the  poles found in the coupled PB-VB systems  ($g_{KR}$ =6) in the isospin 1 configuration. } \label{iso1_2}
\begin{ruledtabular}
\begin{tabular}{lrrrrrrr}
$M_R - i\Gamma/2$ (MeV) $\longrightarrow$ & $1426 -  i143$  &  $1439 -  i194$  &  $2372 - i162$ \\
\hline
Channels $\downarrow$& \multicolumn{3}{c}{Couplings ($g^i$) of the poles to the different channels} \\
\hline
$\bar{K} N$               &$-1.0 - i 0.7$&   $ 0.7 + i 0.3$   &$-0.1 + i 0.0$\\
$\pi \Sigma$             &$ 1.7 + i 1.8$&  $ 1.2 + i 1.3$    &$-0.5 + i 0.2$ \\
$\pi \Lambda$          &$ 0.1 - i 0.3$&  $-1.0 - i 0.6$   &$ 0.3 - i 0.2$ \\
$\eta \Sigma$           &$ 0.1 - i 0.1$&  $-0.7 - i 0.3$   &$ 0.3 - i 0.2$ \\
$K \Xi$                       &$ 0.7 + i 0.8$& $ 1.4 + i 1.0$    &$-0.8 + i 0.0$ \\
$\bar{K^*}N$             &$ 1.7 - i 0.2$&   $ 2.6 - i 0.1$  & $-0.1 - i 0.1$\\
$\rho  \Lambda$       &$-5.6 + i 0.5$&    $-5.8 + i 1.1$   &$ 0.2 + i 0.5$\\
$\rho  \Sigma$          &$ 4.5 - i 0.2$&    $ 6.5 - i 1.4$   &$-0.1 - i 0.8$\\
$\omega  \Sigma$   &$-0.1 + i 0.3$&   $ 0.2 + i 0.1$    &$ 0.0 + i 0.2$\\
$K^* \Xi$                    &$ 5.7 - i 1.1$&    $ 5.0 - i 1.1$  &$ 0.4 - i 0.9$\\
$\phi  \Sigma$           &$ 0.2 - i 0.5$&   $-0.3 - i 0.2$  &$ 0.1 - i 0.3$\\
\end{tabular}
\end{ruledtabular}
\end{table}
The poles at $1426 -  i143$ MeV and  $1439 -  i194$ MeV couple most to $K^* \Xi$ and $\rho \Sigma$ channels, respectively. However these channels are closed for the decay of the corresponding resonances.  The open channels which couple strongly to the pole at $1426 -  i143$ MeV are $\bar{K} N$ and $\pi \Sigma$ and those which couple most to the
 $1439 -  i194$ MeV state are $K \Xi$ and $\pi \Sigma$. These two poles could be associated with the $\Sigma (1480)$ resonance and hope that the information given in the present article will be useful in a better understanding of  this state.

 The pole at $2372 - i162$ MeV is found to couple most to the $K^* \Xi$ channel, however we do not see a clear corresponding peak structure in the 
 $K^* \Xi$ amplitude (shown in Fig.~\ref{fig_iso1}), which might be due to the presence of a negative interference of the pole with the background.
A very little information is available  about the  $\Sigma$ states above 2~GeV region. Thus, for the moment, we do not relate our  state with any known resonance.

\section{Summary}
The work discussed in this manuscript can be summarized as follows:
\begin{enumerate}
\item{One of the important finding of our study is related to the calculation of the loops. We find that the dimensional regularization method can only
be used in the energy region very close to the threshold of the  involved systems. Thus, the cut-off method  is to be used at energies much below the threshold of a system.  To go to higher energies, a Gaussian form factor can be used.  }
\item{Coupling VB to the PB  systems with strangeness $-1$ and isospin 0 reveals large coupling of the low lying $\Lambda$ resonances to the closed VB channels, although the pole positions and the couplings of these resonances to the PB channels remain almost unaltered.  It is important to mention here that the large coupling of the low-lying $\Lambda$'s to the VB channels found in our work do not imply the presence of a large fraction of VB component in the wave function of these resonances since the large mass difference between the two would suppress it. Therefore, the interpretation of the low-lying $\Lambda$'s as PB molecular states does not change. However, our findings could
have some implications on, for example, the photoproduction of the $\Lambda$ resonances where the production mechanism proceeding through exchange of a  vector meson could become important \cite{Nam:2008jy}. This should be verified in future. }
\item{The isoscalar states with higher masses, which have earlier been found to get generated in VB systems, do not get much wider by coupling the open PB channels.   }
\item{In the isospin 1 case, a double pole structure is found near 1430 MeV and we relate this to  $\Sigma (1480)$.}
\end{enumerate}

\section{Acknowledgements}
 This work is partly  supported  by the Grant-in-Aid for Scientific Research on Priority Areas titled ÒElucidation of New Hadrons with
a Variety of Flavors" (E01: 21105006 for K.P.K and A.H) and (22105510 for H.~N) and the authors acknowledge the same. A.~M.~T  
is thankful to the support from the Grant-in-Aid for the Global COE Program ÒThe Next
Generation of Physics, Spun from Universality and EmergenceÓ from the Ministry of Education,
Culture, Sports, Science and Technology (MEXT) of Japan.

\end{document}